\documentclass[aps,prx,reprint,superscriptaddress,twocolumn,footinbib, longbibliography]{revtex4-2}

\pdfoutput=1

\usepackage{amsmath,amsfonts,amssymb,amsthm,mathtools,mathrsfs}
\usepackage{color}
\usepackage{soul}
\usepackage{subfigure}
\usepackage{physics}
\usepackage{dsfont}
\usepackage{footmisc}
\usepackage{hyperref}
\usepackage{url}
\usepackage{textcomp}
\usepackage{rotating}
\usepackage{lmodern}
\usepackage[dvipsnames]{xcolor}
\usepackage{array}
\usepackage{soul}

\definecolor{emerald}{rgb}{0.31, 0.78, 0.47}
\definecolor{blue(ncs)}{rgb}{0.0, 0.53, 0.74}

\hypersetup{
     colorlinks=true,
     linkcolor=magenta,
     filecolor=blue,
     citecolor=emerald,      
     urlcolor =blue(ncs),
}

\DeclareMathAlphabet{\pazocal}{OMS}{zplm}{m}{n}

\newcommand{\BoxH}{\includegraphics[trim=0 12pt 0 0, height=1.0em]{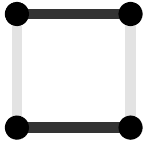}}
\newcommand{\BoxV}{\includegraphics[trim=0 12pt 0 0, height=1.0em]{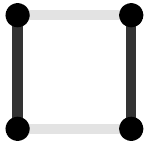}}
\newcommand{\Vertia}{\includegraphics[trim=0 12pt 0 0, height=0.9em]{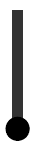}}
\newcommand{\Horiza}{\includegraphics[trim=0 0 0 0, height=0.3em]{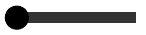}}
\newcommand{\Vertib}{\includegraphics[trim=0 12pt 0 0, height=0.9em]{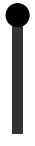}}
\newcommand{\VertiaL}{\includegraphics[trim=0 12pt 0 0, height=0.9em]{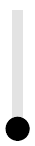}}
\newcommand{\Horizb}{\includegraphics[trim=0 0 0 0, height=0.3em]{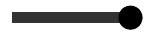}}

\newcommand{\BoxHsmall}{\includegraphics[trim=0 10pt 0 0, height=0.8em]{BoxH.pdf}}
\newcommand{\BoxVsmall}{\includegraphics[trim=0 10pt 0 0, height=0.8em]{BoxV.pdf}}

\newcommand{\Eastoa}{\includegraphics[trim=0 12pt 0 0, height=1.0em]{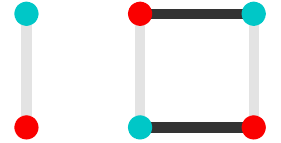}}
\newcommand{\Eastob}{\includegraphics[trim=0 12pt 0 0, height=1.0em]{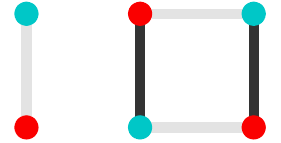}}
\newcommand{\Eastea}{\includegraphics[trim=0 12pt 0 0, height=1.0em]{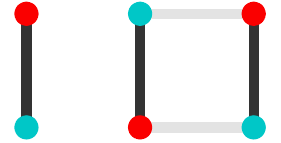}}
\newcommand{\Easteb}{\includegraphics[trim=0 12pt 0 0, height=1.0em]{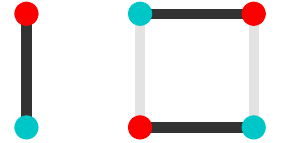}}

\newcommand{\Westoa}{\includegraphics[trim=0 12pt 0 0, height=1.0em]{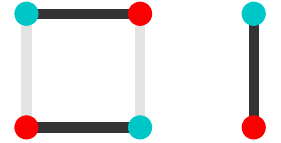}}
\newcommand{\Westob}{\includegraphics[trim=0 12pt 0 0, height=1.0em]{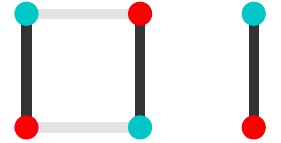}}
\newcommand{\Westea}{\includegraphics[trim=0 12pt 0 0, height=1.0em]{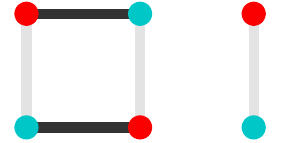}}
\newcommand{\Westeb}{\includegraphics[trim=0 12pt 0 0, height=1.0em]{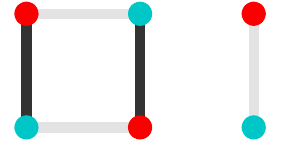}}

\newcommand{\Northoa}{\includegraphics[trim=0 30pt 0 0, height=1.5em]{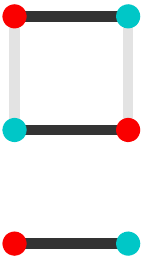}}
\newcommand{\Northob}{\includegraphics[trim=0 30pt 0 0, height=1.5em]{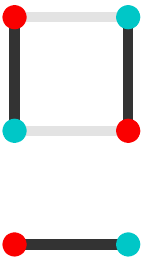}}
\newcommand{\Northea}{\includegraphics[trim=0 30pt 0 0, height=1.5em]{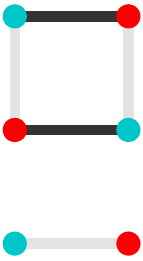}}
\newcommand{\Northeb}{\includegraphics[trim=0 30pt 0 0, height=1.5em]{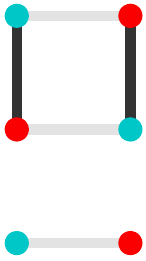}}

\newcommand{\Southoa}{\includegraphics[trim=0 30pt 0 0, height=1.5em]{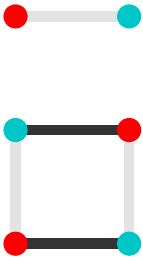}}
\newcommand{\Southob}{\includegraphics[trim=0 30pt 0 0, height=1.5em]{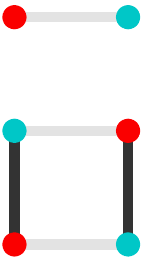}}
\newcommand{\Southea}{\includegraphics[trim=0 30pt 0 0, height=1.5em]{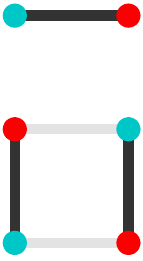}}
\newcommand{\Southeb}{\includegraphics[trim=0 30pt 0 0, height=1.5em]{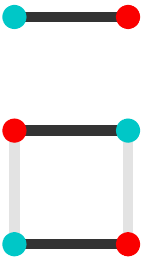}}

\begin{document}

\title{The frustration-free fully packed loop model}

\author{Zhao Zhang}
\email{zhzhang@sissa.it}
\affiliation{SISSA and INFN, Sezione di Trieste, via Bonomea 265, I-34136, Trieste, Italy}
\affiliation{Nordita, KTH Royal Institute of Technology and Stockholm University, Hannes Alfv\'{e}ns v\"{a}g 12, SE-106 91 Stockholm, Sweden}
  
\author{Henrik Schou R{\o}ising}
\email{henrik.roising@nbi.ku.dk}
\affiliation{Nordita, KTH Royal Institute of Technology and Stockholm University, Hannes Alfv\'{e}ns v\"{a}g 12, SE-106 91 Stockholm, Sweden}
\affiliation{Niels Bohr Institute, University of Copenhagen, Jagtvej 128, DK-2200 Copenhagen, Denmark}

\date{\today}

\begin{abstract}
We consider a quantum fully packed loop model on the square lattice with a frustration-free projector Hamiltonian and ring-exchange interactions acting on plaquettes. A boundary Hamiltonian is added to favour domain-wall boundary conditions and link ground state properties to the combinatorics and six-vertex model literature. We discuss how the boundary term fractures the Hilbert space into Krylov subspaces, and we prove that the Hamiltonian is ergodic within each subspace, leading to a series of energy-equidistant exact eigenstates in the lower end of the spectrum. Among them we systematically classify both finitely entangled eigenstates and product eigenstates. Using a recursion relation for enumerating half-plane configurations, we compute numerically the exact entanglement entropy of the ground state, confirming area law scaling. Finally, the spectrum is shown to be gapless in the thermodynamic limit with a trial state constructed by adding a twist to the ground state superposition.
\end{abstract}

\maketitle

%
%%
%%%
\section{Introduction}
\label{sec:Intro}
%%%
%%
%
Ergodicity and its breaking lies at the foundation of modern statistical mechanics. It plays a key role in understanding of why the long-time average of an observable of a single system can be well-approximated by a statistical ensemble average. In quantum systems, any initial state being thermalized necessarily requires each eigenstate of the Hamiltonian to be thermalized, leading to the so-called eigenstate thermalization hypothesis (ETH)~\cite{Deutch91, Srednicki94}. Over the past few years, ETH violation has been realized outside the scope of the integrability and many-body localization paradigms~\cite{ETH, AbaninEA19}, such as in quantum many-body scars (QMBSs) and Hilbert space fragmentation~\cite{AbaninEA19, BernevigEA21}.

Two types of models have played important roles in understanding these novel mechanisms of weak ergodicity-breaking. The first type is kinetically constrained models, which can arise as low energy effective models through a Schrieffer--Wolf transformation~\cite{SchriefferEA66, BravyiEA11}. The dimensionality of constrained Hilbert spaces typically grows as an integer sequence, reflecting an underlying combinatorial structure. In one dimension, a prime example is the PXP model~\cite{LesanovskyEA12, FendleyEA04,TurnerEA18}, which successfully explains Rydberg blockade experiments~\cite{BernienEA17}. The dimensionality of its Hilbert space grows as the Fibonacci numbers with the asymptotic scaling $1.618^N$. In two dimensions, arguably one of the most studied models in classical statistical mechanics and combinatorics is the six-vertex model. With periodic boundary conditions, its Hilbert space dimension grows as $1.540^{N^2}$, following from Lieb's solution to the square ice problem at the ice point for which the weights of the six vertices are identical~\cite{Lieb67}. Sophisticated results have been established when the model is subject to domain-wall boundary conditions (DWBCs), for which a bijection between configurations and alternating sign matrices (ASMs)~\cite{Kuperberg97} has been proven. The exact enumeration of ASMs is a celebrated result in combinatorics~\cite{MillsEA83, Zeilberger96, Kuperberg97, bressoud99}. Notably, progress has also been made in technologies and ideas for realizing classical and quantum spin ice models~\cite{SchifferEA21}, with platforms ranging from arrays of ferromagnetic islands~\cite{WangEA06} to two-dimensional Rydberg atom arrays~\cite{WieseEA13, GlaetzleEA14, CeliEA20, SemeghiniEA21, SamajdarEA21}.

A second type of models studied in weak ergodicity-breaking are frustration-free (FF) Hamiltonians, which have a unique ground state being the superposition of configurations from a usually classical statistical mechanical or combinatorial ensemble. Such Hamiltonians are called frustration-free because their ground state is the simultaneous lowest energy eigenstate of all its local terms. Examples of frustration-free models in 1D include the Motzkin~\cite{BravyiEA12} and the Fredkin spin chain~\cite{SalbergerKorepin16, Luca19}, for which ground state configurations reassemble combinatorial structures known as Motzkin and Dyck walks. Recently, it has been shown that by flipping the signs of some of the projectors, the FF eigenstate can be relocated to the middle of the spectrum, making it qualified for a quantum many-body scar. These systems, as well as the original models, also exhibit Hilbert space fragmentation~\cite{LanglettEA21, RichterEA22, KhudorozhkovEA21}, which refers to the emergence of exponentially many dynamically disconnected subspaces. A classification further distinguishes genuine from ``local'' fragmentation, as related to the scaling of the Krylov subspaces with system size~\cite{SanjayEA22, Berislav22}.

Entanglement entropy (EE) plays an important role in the study of both types of aforementioned models of novel weak ergodicity-breaking. In the first type of models, the growth of EE is used to characterize the slow thermalization behavior of the so-called scarred initial state. In the second type of models in one dimension, the ground state provides an example of violation of area law. Area law here refers to a ground state EE scaling of $S \sim N^{d-1}$ where $d$ is the spatial dimension. A milestone in the study of EE has been Hastings' proof of area law in gapped one-dimensional (1D) systems~\cite{Hastings07}. Recently, a similar result in two dimensions (2D) has been proven for frustration-free models~\cite{AnshuEA21, EisertEA10}. The EE of free fermions generally violate area law, but only logarithmically, $N^{d-1}\ln{N}$~\cite{GioevEA06}. In 1D there are multiple mechanisms that generate beyond logarithmic violation of area law, such as enlarging the degrees of freedom or the local Hilbert space~\cite{ZhangEA22_3}, and strong inhomogeneities~\cite{Vitagliano_2010}. A combination of these approaches also generalizes extensive entanglement growth to Hausdorff dimension one lattices embedded in higher dimensions~\cite{zhang2023entanglement}. One route to generalize area law violation to higher dimensions is offered by the Motzkin and Fredkin spin chains, which are both translationally invariant and can violate the area law strongly, with up to volume-law scaling~\cite{MovassaghEA16, salberger2018fredkin, ZhangEA17, Salberger2EA17, ZhangKlich17}. A crucial ingredient of the Motzkin and Fredkin models is that they allow a height representation that can carry non-local information of the interactions. The first obstacle in generalizing this to 2D is to find a well-defined height function that does not give rise to any ambiguity when going around a closed loop in the lattice. This problem is intrinsically avoided in the context of dimer and fully packed loop (FPL) models~\cite{Henley97, JacobsenEA92, ArdonneEA04, HerdmanEA13}. To further enforce the incremental height change between adjacent plaquettes to be $\pm 1$ on square lattice, we opt for FPLs.

In this manuscript, we combine the two mechanisms of weak ergodicity-breaking in a single model in two dimensions and explore its various features, including Hilbert space fragmentation, ground state entanglement entropy, and an upper-bound on the spectral gap. We find that the notion of a height function alone in two dimensions is not enough to generate beyond area-law EE, due to the strong constraint of gauge invariance. A further decoration of the model in the current manuscript, with enlarged local degrees of freedom and next-nearest neighbor interactions, was recently proposed in two models possessing up to volume EE scaling of ground state ~\cite{ZhangEA22, ZhangEA22_2}. In this manuscript we lay the foundation that make these two generalizations possible, including detailed discussions of ergodicity breaking and its dynamical consequences, why a height model alone is not enough to break area-law (contrary to the 1D case with Motzkin and Fredkin spin chains), and a proof of gaplessness in the thermodynamic limit. The current manuscript also complements the models of Refs.~\onlinecite{ZhangEA22, ZhangEA22_2}, which are more focused on EE area-law violation and phase transitions, from considerations of the global structure of the Hilbert space and excited states.

The paper is organised as follows. In Sec.~\ref{sec:Model}, we introduce the Hilbert space and Hamiltonian of our model. In Sec.~\ref{sec:HilbertSpaceFragmentation} we explore ergodicity breaking in detail (Sec.~\ref{sec:Fragmentation}), in addition to constructing a series of exact eigenstates (Sec.~\ref{sec:Eigenstates}). In Sec.~\ref{sec:Entropy} we compute numerically the exact ground state bipartite entanglement entropy between half-systems, yielding area law scaling. In Sec.~\ref{sec:Bulkgap} we construct a trial state with vanishing energy in the thermodynamic limit, demonstrating that the model is gapless. Finally, we provide a summary and propose future directions. In Appendix~\ref{sec:Ergodicity} we prove that FPL configurations with a fixed boundary are ergodic with respect to plaquette flipping in the bulk. In Appendix~\ref{sec:sufficient} we provide a sufficiency proof for exact unentangled exited states. In Appendix~\ref{sec:Recursive} we explain how half-system configurations can be counted recursively, which is used to calculate the entanglement entropy, and in Appendix~\ref{sec:MonteCarlo} we devise a Monte Carlo algorithm to faithfully sample the classical configuration space. Finally, in Appendix~\ref{sec:Generalizations} we provide an example of a 2D frustration-free FPL model that does not suffer from boundary constraints.

%
%%
%%%
\section{The model and its dual lattice formulation}
\label{sec:Model}
\begin{figure}[t!bh]
	\centering
	\includegraphics[width=\linewidth]{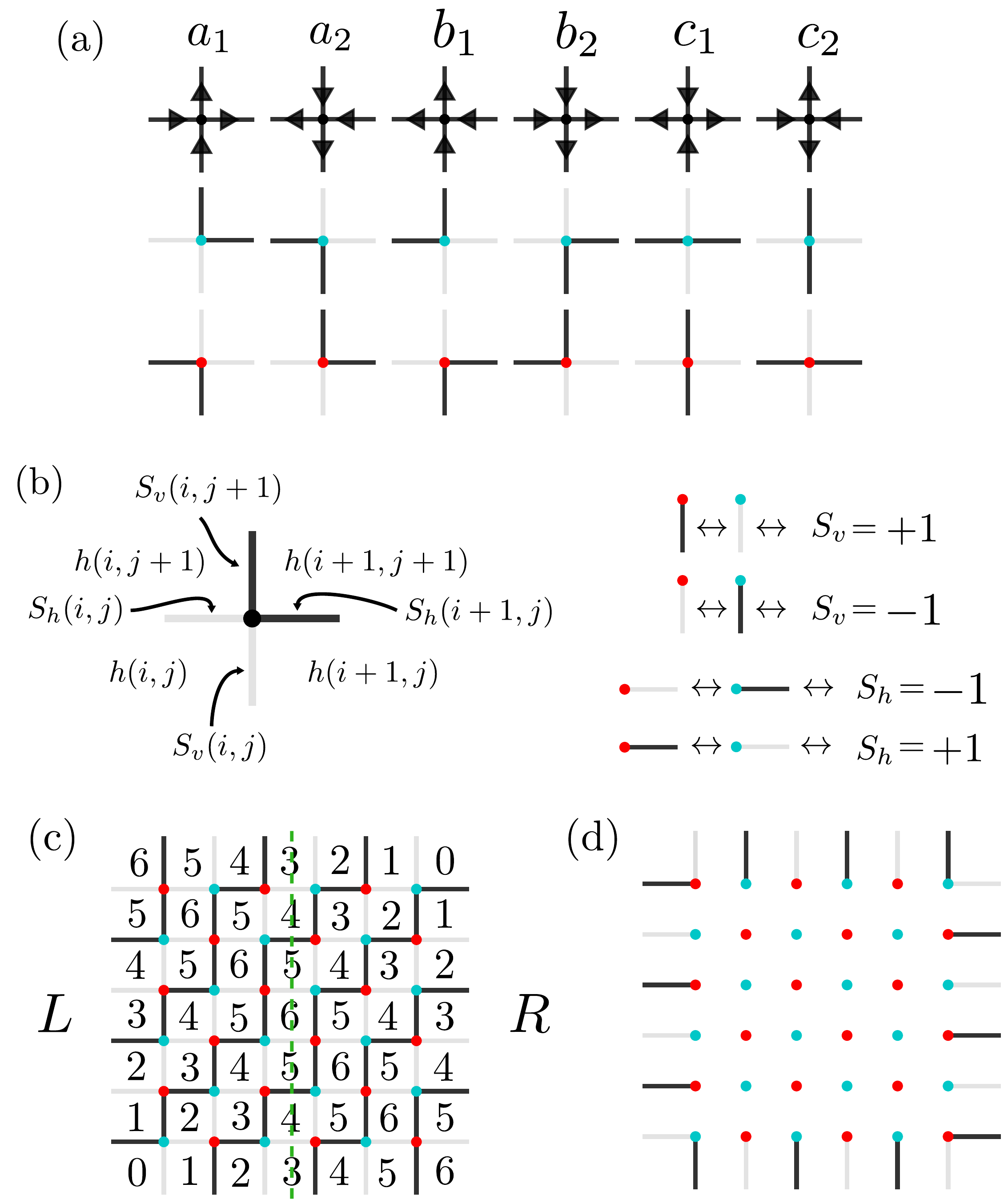} 
	\caption{(a) Mapping between configurations of the six-vertex model (top row) and fully packed loop configurations (FPLs, lower two rows), with rules alternating on the even (cyan dots)  and odd (red dots) sublattices~\cite{Justin09}. (b) Height (on plaquettes) and spin (on bonds) representation of FPLs, which are related as $\vec{S} = (S_v,~S_h) = \vec{\nabla} h$ with discrete derivatives, see the main text. The FPL constraint amounts to imposing $\vec{\nabla} \times \vec{S} = 0$ around each vertex. (c) Maximal height configuration for $N = 6$ with domain-wall boundary condition DWBC1 in the height representation. The dashed green line represents the cut between the left ($L$) and right ($R$) half-systems of which we calculate the ground state entanglement entropy in Sec.~\ref{sec:Entropy}. (d) Boundary conditions labelled DWBC2 relevant for the unique and exact excited state with energy $E = 4N$ described in Sec.~\ref{sec:Eigenstates}.}
	\label{fig:HeightRep}
\end{figure}

%%%
%%
%
We consider a square lattice $G$ of $N \times N$ vertices, with binary degrees of freedom living on the horizontal and vertical bonds connecting neighbouring vertices. We constrain the Hilbert space to that of the fully packed loop (or 2-dimer) coverings of the lattice by the diagonal Hamiltonian $H_0 = V\sum_{i,j=1}^N \big(\sigma_v(i,j)+\sigma_h(i,j)+\sigma_v(i,j+1)+\sigma_h(i+1,j)\big)^2$, where subscripts $v$ and $h$ refer to vertical and horizontal bonds respectively, and $\sigma=+1$ ($\sigma = -1$) for a covered (uncovered) bond. In the limit $V \gg 1$ we can effectively work in the ground state manifold of $H_0$.
\begin{figure}[t!bh]
	\centering
	\includegraphics[width=\linewidth]{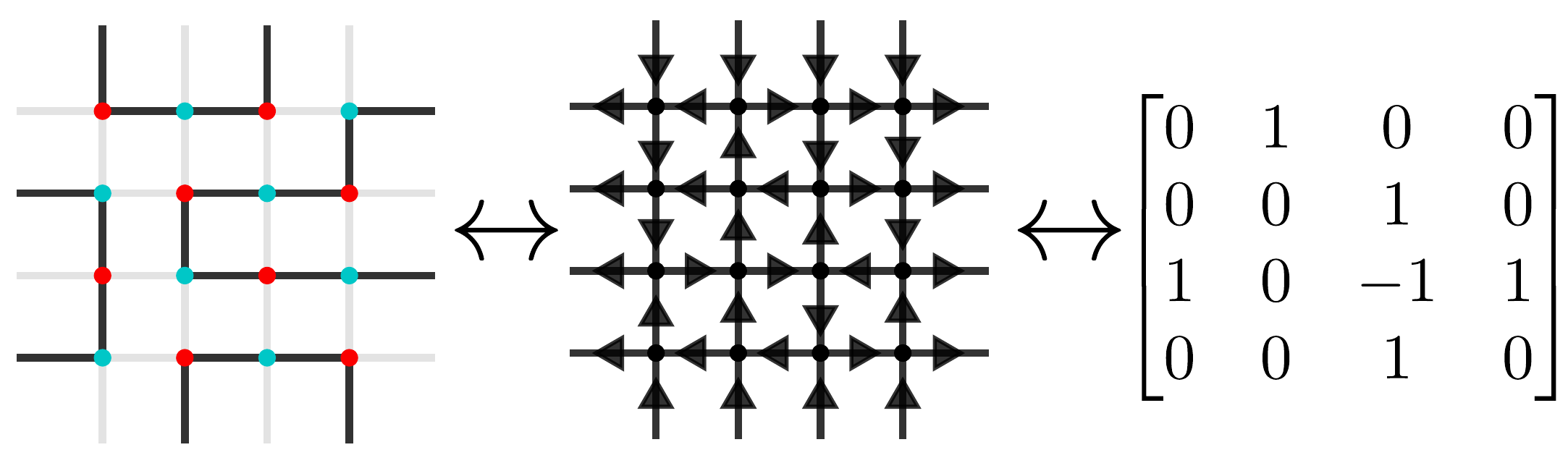} 
	\caption{Bijections with domain-wall boundary conditions. Left: fully packed loop configuration, middle: six-vertex model configuration, right: alternating sign matrix. The mapping between the first two representations is explained in Fig.~\ref{fig:HeightRep} (a), and the mapping between the latter two amounts to assigning $a$ and $b$ vertices to $0$, and $c$ vertices to $\pm 1$~\cite{Justin09}.}
	\label{fig:bijections}
\end{figure}
In the constrained Hilbert space we consider the low-energy effective Hamiltonian
\begin{equation}
\begin{aligned}
H &= H_{\mathrm{bulk}}+H_{\partial}=\sum_{p \in \mathrm{bulk}} P_{p} + H_{\partial}, \\
P_{p} &= \frac12 \left( \ket{\BoxV} - \ket{\BoxH} \right) \left( \bra{\BoxV} - \bra{\BoxH} \right)_p.
\label{eq:RKmodel}
\end{aligned}
\end{equation}
The Rokhsar--Kivelson projectors, $P_p$, contain the diagonal potential term $\frac{1}{2}(\ket{\BoxVsmall}\bra{\BoxVsmall}+\ket{\BoxHsmall}\bra{\BoxHsmall})$ and the off-diagonal ring-exchange kinetic term $-\frac{1}{2}(\ket{\BoxVsmall}\bra{\BoxHsmall}+\ket{\BoxHsmall}\bra{\BoxVsmall})$ that allows parallel covered bonds to resonate~\cite{RokhsarKivelson88}. We will refer to plaquettes that contain two parallel covered bonds as \emph{flippable}, and \emph{unflippable} otherwise. Here bonds are either covered (black) or uncovered (light gray), with the bond-spin conversion rules in Fig.~\ref{fig:HeightRep}~(b). The sum above runs over the bulk plaquettes, which there are $(N-1)^2$ of. The boundary terms
\begin{equation}
\begin{aligned}
H_{\partial} &= H_{\partial}^{t} + H_{\partial}^{b} + H_{\partial}^{l} + H_{\partial}^{r} + 2N, \\
H_{\partial}^{t} &= \sum_{x = 1}^N (-1)^x \ket{\Vertia}\bra{\Vertia}_{x,N}, \\
H_{\partial}^{b} &= \sum_{x = 1}^N (-1)^{x+1} \ket{\Vertib}\bra{\Vertib}_{x,1} \\
H_{\partial}^{l} &= \sum_{y = 1}^N (-1)^{y} \ket{\Horizb}\bra{\Horizb}_{1,y}, \\
H_{\partial}^{r} &= \sum_{y = 1}^N (-1)^{y+1} \ket{\Horiza}\bra{\Horiza}_{N,y}. \\
\end{aligned}
\label{eq:BoundaryTerm}
\end{equation}
impose a domain-wall boundary condition (labelled DWBC1) on the ground state, where every other bond is covered along the boundary (see Fig.~\ref{fig:HeightRep} (c)). Above, subscripts denote the (horizontal or vertical) bond position $(x, y)$, as counted from the lower left of the graph with row-major ordering. With these boundary conditions FPL coverings are also in bijection to alternating sign matrices~\footnote{Alternating sign matrices are matrices with elements $0$, $-1$, or $+1$ such that the sum of each column and row is $1$, and where the $+1$ and $-1$ elements alternate along rows and columns. The mapping to six-vertex model configurations is obtained by assigning $c_1$ vertices $+1$, $c_2$ vertices $-1$, and all the remaining four vertices $0$~\cite{bressoud99, Kuperberg97}.}, see Fig.~\ref{fig:bijections}.

The bulk Hamiltonian has the apparent $\mathbb{Z}_2$ symmetry $\mathcal{R}$ of reversing the covering of all the bonds, satisfying
\begin{equation}
    [H_{\mathrm{bulk}},\mathcal{R}]=0,
    \label{eq:bulksymmetry}
\end{equation}
which is broken by the boundary terms:
\begin{equation}
    \{H_{\partial}, \mathcal{R}\}=4N\mathcal{R}.
    \label{eq:bondflip}
\end{equation}
The commutation relation of Eq.~\eqref{eq:bulksymmetry} can be understood by observing that all the projectors $P_p$ are invariant upon interchanging covered and uncovered bonds. The anti-commutation relation of Eq.~\eqref{eq:bondflip} becomes apparent by noticing that $\ket{\Vertia}\bra{\Vertia}\coloneqq 1-\ket{\VertiaL}\bra{\VertiaL} = \frac{1}{2}(1+\ket{\Vertia}\bra{\Vertia}-\ket{\VertiaL}\bra{\VertiaL})$, which makes $H_\partial-2N$ anticommute with $\mathcal{R}$.
The full Hamiltonian also has a hidden symmetry given by Wieland's gyration~\cite{gyration}, which reverses the coverings around only the non-flippable plaquettes, while leaving the flippable ones unchanged. 

\begin{figure}[t!bh]
	\centering
	\includegraphics[width=0.6\linewidth]{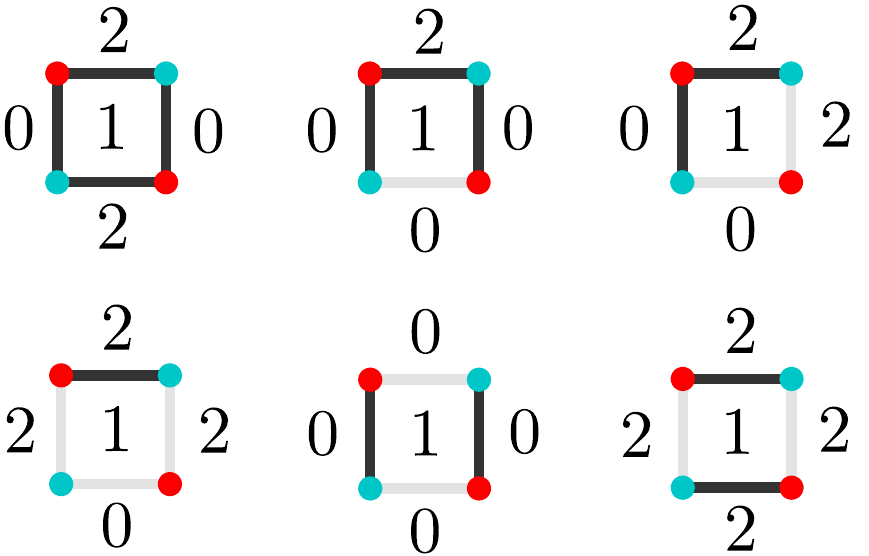} 
	\caption{Flippable plaquettes are precisely those for which the height is above or below all its four neighbours.}
	\label{fig:Flippable}
\end{figure}

The off-diagonal terms in Eq.~\eqref{eq:RKmodel} relate FPL configurations differing on a single plaquette with parallel bonds covered in different directions. This is conveniently expressed in the height representation on the dual lattice~\cite{Henley97, ArdonneEA04}. On the square lattice the height field is integer-valued and changes in units of $\pm 1$ between neighbouring plaquettes. Using the bipartite rules summarized in Fig.~\ref{fig:HeightRep} (b) the conversion between spins and height is compactly expressed as $\vec{S} = (S_v,~S_h) = \vec{\nabla} h$, with $\partial_x h(i,j) \coloneqq h(i+1,j)-h(i,j)$ and $\partial_y h(i,j) \coloneqq h(i,j+1)-h(i,j)$. The FPL constraint then amounts to imposing $\vec{\nabla} \times \vec{S}(i,j) = 0$ around each vertex. It can easily be verified that these definitions make precisely the (flippable) plaquettes with two parallel covered bonds local extrema in the height, see Fig.~\ref{fig:Flippable}.

On the dual lattice, the model can be expressed as the kinetically constrained Hamiltonian
\begin{equation}
\begin{aligned}
    H^*
    =\sum_{p\in \text{bulk}}&\big(\Pi_{p}^>+\Pi_{p}^< -\Pi_{p}^> h_p^+ \Pi_p^<- \Pi_{p}^<h_p^-\Pi_{p}^> \big) +H^*_{\partial},
\end{aligned}
\label{eq:Dual}
\end{equation}
where $\Pi_p^{>(<)}$ projects onto the state where the four neighboring plaquettes of $p$ all have the same height above (resp.~below) the height of $p$, and $h_p^{+(-)}$ increases (resp.~decreases) the height of plaquette $p$ by $2$. The boundary term is given by
\begin{equation}
\begin{aligned}
    H_{\partial}^{\ast}
    &= h(1,1)-h(1,N+1) \\
   &\hspace{10pt}-h(N+1,1)+h(N+1,N+1)+2N.
\end{aligned}
\label{eq:DualBC}
\end{equation}
The height representation on the dual lattice reveals another symmetry of the Hamiltonian. The height difference along each row and column (or along any other path connecting two boundary plaquettes), which are equal to the sum of spins $\sum_{j=1}^{N}S_v(i,j)$ and $\sum_{i=1}^{N}S_h(i,j)$, respectively, is conserved. In fact, in the absence of kinetic (off-diagonal) terms on the boundary, which transform one boundary configuration into another~\footnote{One example of such kinetic terms is an on-site operator that creates or annihilates a covered bond, another example is the swapping of a neighbouring pair of covered and uncovered bonds.}, there are a linear number of local discrete symmetries:
\begin{equation}
\begin{aligned}
     [H, S_h(1,j)] &= 0 \quad & \forall j=1,N+1, \\
     [H, S_v(i,N+1)] &= 0 \quad & \forall i=1,N+1, \\
     [H, S_h(N+1,j)] &= 0 \quad & \forall j=1,N+1, \\
     [H, S_v(i,1)] &= 0 \quad & \forall i=1,N+1,
 \end{aligned}
 \label{eq:LocalSymmetries}
 \end{equation}
which are responsible for the ergodicity breaking to be discussed in the next section. This makes the model possess ``local fragmentation'' in the terminology of Ref.~\cite{Berislav22}, to be contrasted with ergodicity breaking due to either discrete or continuous global symmetry such as the total magnetization in spin chains, or genuine fragmentation in models such as caused by pair flipping~\cite{SanjayEA22}. In Appendix~\ref{sec:Ergodicity} we prove that the bulk Hamiltonian is ergodic within each Krylov subspace spanned by all the FPL configurations sharing the same boundary configuration.

%
%%
%%%
\section{Ergodicity breaking and exact eigenstates}
\label{sec:HilbertSpaceFragmentation}
%%%
%%
% 
In this section we identify exact unentangled and finitely entangled eigenstates, by making simple observations of constraints on the boundary configurations. Among the exact eigenstates we construct the exact and unique ground state and its twin ceiling state.

\subsection{Product eigenstates and finitely entangled eigenstates}
\label{sec:Fragmentation}
\begin{figure}[t!bh]
	\centering
	\includegraphics[width=\linewidth]{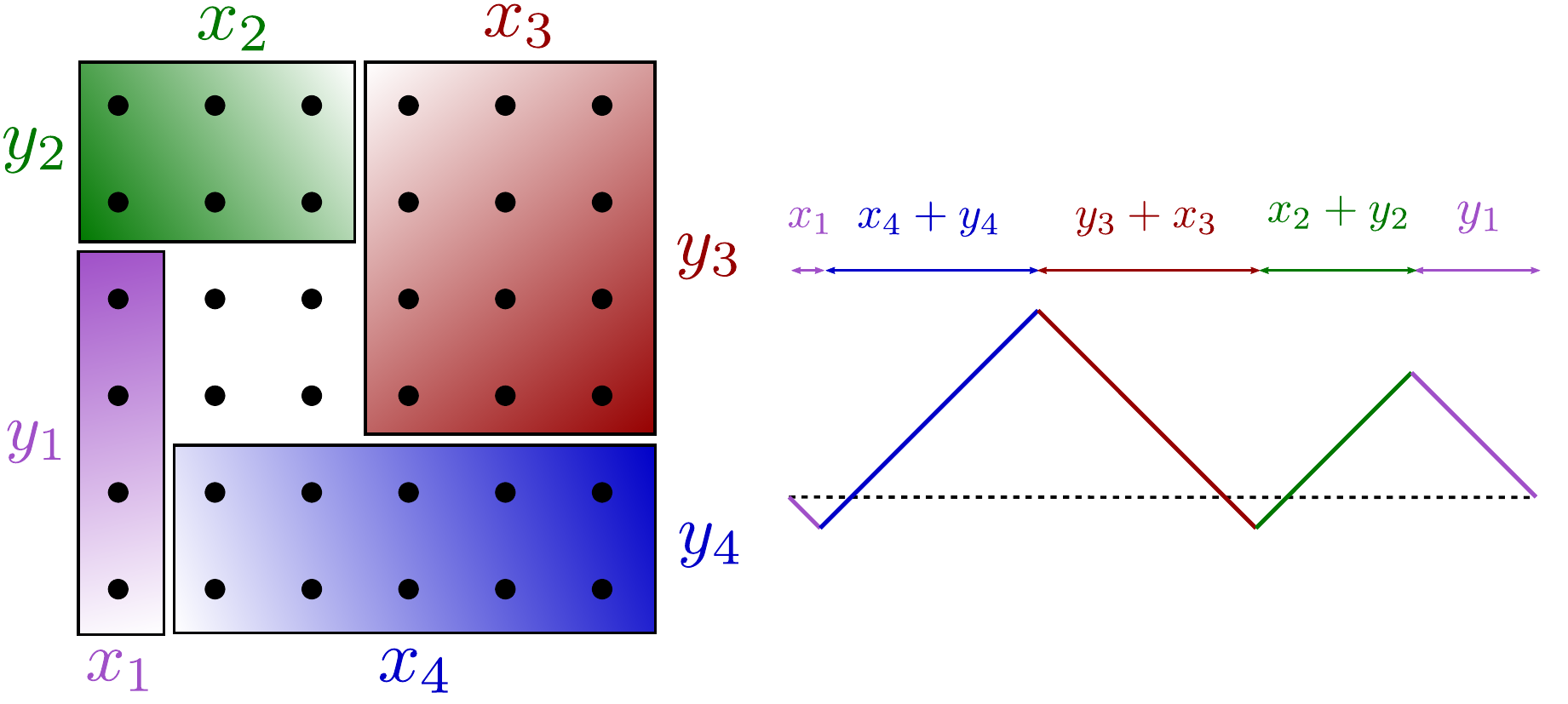}
	\caption{A boundary configuration with four monotonic segments (two kinks and two anti-kinks), demanding four rectangular monotonic regions near four corners, leaving a core with a single flippable plaquette. The right panel shows the corresponding height going around the lattice boundary counter-clockwise.}
	\label{fig:FourKink}
\end{figure}

If we follow the height counter-clockwise around the perimeter, the fact that it must come back to the same value after a full cycle implies that there must be at least one pair of height kink and anti-kink along the graph perimeter, as the height of neighbouring plaquettes must differ by $\pm 1$. Moreover, if anywhere on the lattice, there is a segment, such as those in the same color in Fig.~\ref{fig:FourKink}, in which the height changes monotonically, and which has a right-angle turn, the height in the entire rectangle spanned by the two perpendicular sides must also change monotonically between diametrically opposing corners of the rectangle. We will henceforward refer to this as the \emph{convexity lemma}, and use it repetitively. By ``segment'' we here refer to any path between two plaquettes with a single right-angle turn, i.e., any ``L''-shaped path. The key point is that since flippable plaquettes correspond to local extrema in the height, a monotonic segment can not contain any flippable plaquettes.

An immediate consequence of the convexity lemma is that a monotonic segment on the perimeter cannot cover two corners. Thus, besides the special case of two kinks being located on diagonally opposing corners as described below, there must be at least two pairs of alternating kink and anti-kink, with the sum of lengths of every other segment being $2N$, since the height along half of the perimeter must increase, and decrease along the remaining half, the net height change along the entire perimeter (of length $4N$) is zero. A four-segment case is depicted in Fig.~\ref{fig:FourKink}.

\begin{figure}[t!bh]
	\centering
	\includegraphics[width=\linewidth]{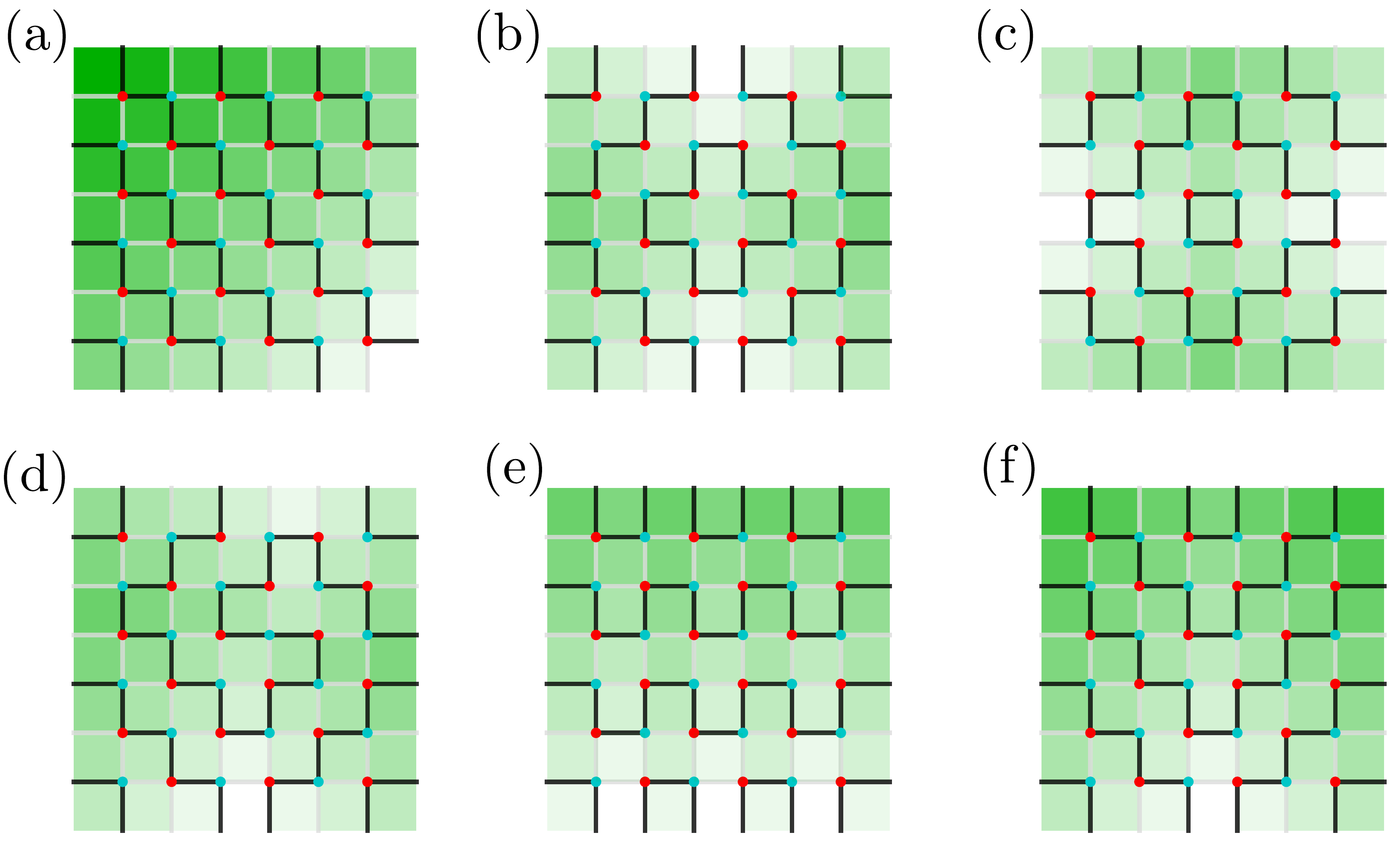} 
	\caption{Representatives of the two types of product eigenstates in isolated one-dimensional Krylov subspaces, with the opaque green color scale reflecting the height. (a), (e), and (f) have straight height gradient lines penetrating to opposite sides of the lattice, while (b), (c), and (d) have a saddle point of height gradients in the bulk. The energy of (a), (b), (c), (e), (f) is always $2N$, while the energy of (d) is $2N - 2$. Up to translations and rotations, the only remaining product states not depicted here have height functions being monotonic in one direction, and globally shifted parallel random walks in the orthogonal direction.}
	\label{fig:ExcitedFPLProduct}
\end{figure}

As for the exceptional kink-antikink case, the height change must be monotonic in each row and column inside the bulk, so there are no flippable plaquettes in this case. We have thus found the first product eigenstate, as depicted in Fig.~\ref{fig:ExcitedFPLProduct}~(a) (for which the kink and the anti-kink are located at diagonally opposing corners). With two kink-antikink pairs, it is still easy to pick boundary configurations that only allow one bulk configuration. First, notice that the only way in which none of the four segments go around a corner is for each of them to cover exactly one side of the lattice, including one of two corner bonds in either end. This gives the ground state boundary configuration depicted in Fig.~\ref{fig:HeightRep}~(c). Otherwise, each segment must go around one corner and one corner only.

If the four rectangles formed by the four monotonic segments cover the entire square lattice, it also results in a product eigenstate. This gives the states shown in Fig.~\ref{fig:ExcitedFPLProduct}~(b), (c), and (d). Together, up to translations and rotations, they exhaust all product eigenstates with two pairs kink-antikink on the boundary. This can be seen by writing each segment length as $l_i=x_i+y_i$ for $i = 1,\dots,4$, see Fig.~\ref{fig:FourKink}. Monotonicity within each rectangular region and alternation of monotonicity between regions require the constraints $\min(x_1+x_3, y_1+y_3)$ and $\min(x_2+x_4, y_2+y_4)$ to be either $N$ or $N-1$ for the excited state to be a product state. The former case gives (b) and (c), up to translations, and the case latter gives (d), up to translations and $\frac{\pi}{2}$ rotations.


By now it can be seen that in general, a rule in searching boundary configurations yielding product eigenstates is that each plaquette in the lattice is penetrated by at least one height gradient line. Here gradient line just means any directed path (among multiple choices) along which the height changes monotonically and should not be confused with the direction of steepest descend. Gradient lines must start and end at boundaries, as loops are not allowed and endpoints imply a flippable plaquette. We now prove that the only possible full coverage of gradient lines, other than the special ones going around four corners as we discussed above, are parallel straight line coverings, corresponding to product states of the type Fig.~\ref{fig:ExcitedFPLProduct}~(e) and (f).

If there is at least one plaquette being traversed by a gradient line that starts and ends on opposite sides, we call this plaquette $(i,j)$, otherwise, it becomes the one of special cases to be discussed in the next paragraph. Assume the plaquette $(i,j)$ is traversed by a gradient line starting from side $a$ and ending on side $b$. Since $a$ and $b$ are on opposite sides, without loss of generality, we can assume the coordinates of the two end points are $(i_a,1)$ and $(i_b,N)$, with $i_a\le i \le i_b$. We then immediately have that the entire region between $i_a$ and $i_b$ is traversed by straight gradient lines from the  bottom to the top of the lattice, by the convexity lemma. But the lines in the region with first coordinate $i \not\in [i_a,~i_b]$ will have no where to go except turning to the adjacent side, resulting in either a curl or violation of the convexity lemma. We thus conclude that if we have a product state in which one plaquette is traversed by a gradient line that starts and stops on opposite sides, then all plaquettes must be traversed by straight and parallel gradient lines starting and ending on opposite sides.

If however, all gradient lines turn to end on adjacent sides, an analogous argument will result in product states of the type in Fig.~\ref{fig:ExcitedFPLProduct} (b), (c) or (d). Hence we have exhausted all the possible unentangled eigenstates. Alternatively, the above argument can be made elegantly in the six-vertex language. As such, a discrete version of electrostatics dictates that gradient lines, coming from strings of unflippable plaquettes, are curl-free and source-free. These are precisely what the FPL configurations shown in Fig.~\ref{fig:ExcitedFPLProduct} correspond to when tracing their height profile.

Apart from product eigenstates, the lowest entangled frustration-free eigenstate corresponding to the boundary configuration in Fig.~\ref{fig:FourKink}, which is a equal superposition of two FPLs differing only by the center plaquette, with EE of $\ln 2$. One can further explicitly diagonalize a $2\times 2$ plaquette system subject to DWBC, which is satisfied by the four-plaquette core surrounded by four monotonic rectangular wings. These states have area law entanglement entropy of the size of the core, which can be anywhere between $1$ and $N$, if the cut is through it. 

\subsection{Exact eigenstates}
\label{sec:Eigenstates}
As we prove in Appendix \ref{sec:Ergodicity}, each consistent boundary configuration forms a Krylov subspace, in which a uniform superposition of all FPLs with that boundary configuration is an eigenstate. Among these would-be degenerate eigenstates, the boundary Hamiltonian picks the one with DWBC1 to be the unique and global ground state
\begin{equation}
\ket{\mathrm{GS}} = \frac{1}{\sqrt{A(N)}} \sum_{\pazocal{F} \in \mathrm{FPL\ with\ DWBC1}} \ket{\pazocal{F}},
\label{eq:RKGroundState}
\end{equation}
where the normalization constant
\begin{equation}
A(N) = \prod_{m = 0}^{N-1} \frac{(3m+1)!}{(N+m)!},
\label{eq:ASM}
\end{equation}
is equal to the number of ASMs~\setcounter{footnote}{19}\footnote{The asymptotics of $A(N)$ is given by $A(N) \sim N^{-5/36} (3\sqrt{3}/4)^{N^2}$~\cite{RobbinsNumbers}.} due to the domain-wall boundary conditions realized by the ground state~\cite{MillsEA83, Zeilberger96, Kuperberg97, bressoud99}, cf.~Fig.~\ref{fig:bijections}.

The Hamiltonian can be deformed locally with a large class of parameters. One can consider the graph Laplacian that gives the Hamiltonian. That graph turns out to have chordless cycles only of length four, originating from flipping two corner-sharing plaquettes in different orders. The projectors acting on each plaquette $p$ can be deformed in a frustration-free manner independently by a spatially varying angle $\theta_p$ not subject to any consistency relation of the type in Ref.~\cite{ZhangEA17}: $P_{p}(\theta_p) = \left( \cos\theta_p\ket{\BoxVsmall} - \sin\theta_p\ket{\BoxHsmall} \right) \left( \cos\theta_p\bra{\BoxVsmall} - \sin\theta_p \bra{\BoxHsmall} \right)_p$. Furthermore, it appears possible to deform the Hamiltonian in a neighborhood-resolved manner such that the ground state is a superposition of weighted six-vertex configurations. The detailed investigation of this quantum six-vertex model beyond the ice point, we intend to undertake in future work.

The necessary and sufficient condition for a boundary configuration allowing an integer energy eigenvalue is most succinctly expressed in the six-vertex language: Firstly, the total number of inward and outward pointing arrows must be the same; secondly, if the lattice is divided along any row or column, the difference between inward and outward pointing arrows on either side of the partition line can not exceed $N$, see Fig.~\ref{fig:BackwardsInduction} in Appendix~\ref{sec:sufficient}, in which the sufficiency condition is proven with backward induction. The necessity follows straightforwardly from the conservation of arrows at each vertex, so any partition of the lattice must leave the interior arrows along the cut capable of balancing the net arrow flow in the exterior.

An eigenstate with $2n$ flipped boundary spins relative to DWBC1 (in Fig.~\ref{fig:HeightRep} (c)), and otherwise being annihilated by the bulk Hamiltonian in a construction similar to the ground state, will have energy $E = 2n$. As explained above and proved in Appendix~\ref{sec:sufficient}, a consistent boundary condition in the height representation requires color balancing, meaning that there is a balanced number of covered boundary bonds connected to vertices of either sublattice, i.e.~$n$ red and $n$ blue vertices connected to covered boundary bonds. The excited state for a consistent boundary condition $\pazocal{S}_n$ (the set of boundary spins) thus takes the form
\begin{equation}
    \ket{2n,\pazocal{S}_n} \propto \sum_{\pazocal{F} \in \mathrm{FPL\ with\ }\pazocal{S}_n} \ket{\pazocal{F}},
\label{eq:ExcitedState}
\end{equation}
where the sum runs over FPLs with boundary condition $\pazocal{S}_n$. A simple estimate for the degeneracy of level $2n$ is $d_{2n} \leq \binom{2N}{n}^2$ since there are $\binom{2N}{n}$ ways in which both $n$ blue and $n$ red boundary spins can be flipped from DWBC1. This is an upper bound since it includes a handful of boundary configurations violating the second part of the sufficiency condition explained above, so that global conservation of arrows is respected, but where the arrow conservation in a subsystem including the partition line is violated~\footnote{For $N=6$ for instance, explicit counting results in a total of $2\;668\;952$ consistent boundary modes, whereas the formula $g(N) \coloneqq \sum_{n=0}^{2N} \binom{2N}{n}^2 = \frac{4^{2N} \Gamma(2N+\frac12)}{\sqrt{\pi}\Gamma(2N+1)}$ predicts $g(6) = 2\;704\; 156$.}. For $n = 0$ we recover the unique ground state with $E = 0$ and $\pazocal{S}_0 = \mathrm{DWBC1}$, and for $n = 2N$ we get the unique excited state with $E = 4N$ and $\pazocal{S}_{2N} = \mathrm{DWBC2}$. This $E = 4N$ state is related to the ground state by the $\mathbb{Z}_2$ operator $\pazocal{R}$ from Eq.~\eqref{eq:bulksymmetry} and~\eqref{eq:bondflip}.

\subsection{Relation to quantum many-body scars}
\label{sec:RelationtoQMBS}

Here we comment on how our mechanism of weak ergodicity-breaking contrasts and complements the known paradigms of spectrum generating algebra~\cite{MarkEA20} and the Shiraishi--Mori embedding formalism~\cite{ShiraishiEA17}. Our Hamiltonian resembles the Shiraishi--Mori formalism, with a target space given by the space of all classical fully packed loops (FPLs). The diagonal boundary Hamiltonian lifts the \emph{a priori} ground state degeneracy, making the ground state unique on the open square grid. Our exact frustration-free excited states are also equidistant in energy. Yet, the majority of our states are not related by ladder operaters, except for the ground state and the ceiling state, which are related by the gyration operator~\cite{gyration}. Furthermore, the energy of our exact excited states are of order $N$ while the spectrum ranges from $0$ to order $N^2$, so the eigenstates do not have finite energy density in the thermodynamic limit and hence do not qualify for QMBS states. The boundary puts stringent constraints on 2D models in constrained Hilbert spaces, to the point that even product state can become eigenstates, which is not observed in 1D, and could lead to potential applications in quantum technology.

\section{Bipartite entanglement entropy}
\label{sec:Entropy}
The bipartite entanglement entropy of the half-system defined by the central cut shown in Fig.~\ref{fig:HeightRep} (c) is given by
\begin{equation}
S = - \sum_{\lbrace \vec{m} \rbrace} p_{\vec{m}} \ln{ p_{\vec{m}} },
\label{eq:Entropy}
\end{equation}
where $\vec{m} = (h(N/2,1), h(N/2,2), \dots, h(N/2,N))$ is the height field across the cut. In what follows we will evaluate this for the ground state using an exact recursion relation.

\subsection{Schmidt decomposition and exact recursion}
\label{sec:SchmidtDecomp}
The numbers $p_{\vec{m}}$ are obtained by Schmidt decomposing the ground state:
\begin{equation}
\ket{\mathrm{GS}} = \sum_{\lbrace \vec{m} \rbrace} \sqrt{p_{\vec{m}}} \ket{\pazocal{P}_{L,\vec{m}}(N/2) } \otimes \ket{\pazocal{P}_{R,\vec{m}}(N/2) },
\label{eq:SchmidtDecomp}
\end{equation}
where $\ket{\pazocal{P}_{L,\vec{m}}(N/2) }$ (resp.~$\ket{\pazocal{P}_{R,\vec{m}}(N/2) }$) is the normalized sum of FPLs in the left (resp.~right) half-system with heights $\vec{m}$ imposed along the right (resp.~left) edge (and DWBC on the remaining three). The number of different height configurations along the (full system) central cut is given by the number paths with height changes of $\pm 1$ across neighbouring plaquettes, starting and ending at the same height of $N/2$:
\begin{equation}
\pazocal{N}_{\vec{m}} = \binom{N}{N/2}.
\label{eq:Nomvecs}
\end{equation}
The coefficient $p_{\vec{m}}$ above then takes the form
\begin{equation}
p_{\vec{m}} = \frac{\lvert \pazocal{P}_{L, \vec{m}}(N/2) \rvert \lvert \pazocal{P}_{R, \vec{m}}(N/2) \rvert}{A(N)},
\label{eq:pmcoeff}
\end{equation}
where $\lvert \pazocal{P}_{R,\vec{m}}(N/2) \rvert$ is the number of FPLs in the right half-system with the left boundary $\vec{m}$. By reflection symmetry around the central cut we have $\lvert \pazocal{P}_{R,\vec{m}}(N/2) \rvert = \lvert \pazocal{P}_{L, \vec{N} -\vec{m} }(N/2) \rvert$ , with $\vec{N} = (N, N, \dots, N)$. Closed-form expressions for $\lvert \pazocal{P}_{L, \vec{m}}(N/2) \rvert$ in terms of contour integrals were given in Ref.~\cite{ColomoEnum}, however, for the purpose of extracting the EE scaling, it is more efficient to use a recurrence relation to enumerate them for incrementally growing $N$. We refer to Appendix~\ref{sec:Recursive} for details on the recursion relation~\footnote{The enumeration of half-system configurations for a subset of the $\vec{m}$'s has been established with available asymptotics, namely those with ``U-turn'' boundaries~\cite{PavelEA17} and some with certain symmetry constraints~\cite{FischerEA19}.}.

\subsection{Scaling of the entanglement entropy}
\label{sec:Evaluation}
The combinatoric nature of the counting problem explained above, related to the rapid growth of $A(N)$~\cite{Note20}, and the exponential slowdown encountered when solving the recursion relation numerically still pose as a practical challenge. 

\begin{figure}[t!bh]
	\centering
	\includegraphics[width=\linewidth]{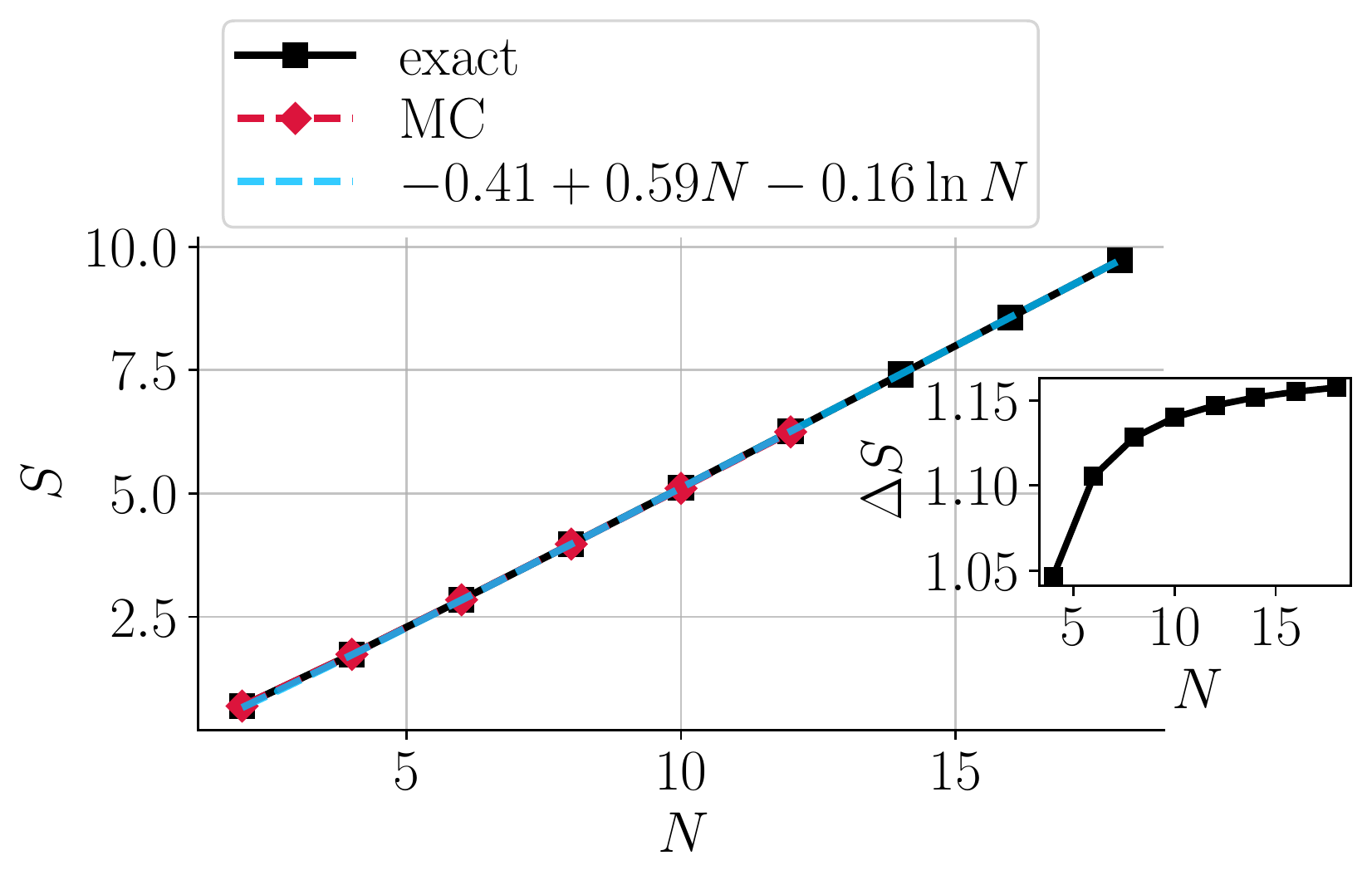} 
	\caption{The bipartite entanglement entropy of the half-system from Eq.~\eqref{eq:Entropy}, as obtained with the exact recursion relation (black squares) described in Appendix~\ref{sec:Recursive} and with the Monte Carlo method (red diamonds) described in Appendix~\ref{sec:MonteCarlo}. The blue dashed line shows a fit to the last three data points. The inset shows the difference between consecutive points, $\Delta S(N) = S(N)-S(N-2)$, which from the upper bound of Eq.~\eqref{eq:EntropyBound} is expected to saturate to a constant for large $N$. }
	\label{fig:Entanglement}
\end{figure}

In Fig.~\ref{fig:Entanglement} we show the outcome of calculating the entanglement entropy of the half-system from the exact recursion relation. A Monte Carlo algorithm, which does not rely on ergodicity and which faithfully samples FPL configurations, was devised and verifies the results with excellent accuracy. Details of the Monte Carlo method are provided in Appendix~\ref{sec:MonteCarlo}. The entanglement entropy resulting from the exact recursion relation has area law scaling. Due to the shape of our lattice (having corners) one generally expects subleading corrections like $\ln{N}$. On the cylinder and torus the general behaviour is a non-universal linear term and a universal constant related to the Lifshitz field theory relevant at the quantum critical point~\cite{ArdonneEA04, HsuEA09, Fradkin13}. 

To shed further light on the scaling we can state a simple upper bound on Eq.~\eqref{eq:Entropy} in the height field representation. The extreme case is obtained for a flat distribution with $p_{\vec{m}} = 1/\pazocal{N}_{\vec{m}}$: 
\begin{equation}
S \le  \ln{ \pazocal{N}_{\vec{m}} } \sim N\ln{2} - \frac12 \ln{N} + \frac12 \ln{\frac{2}{\pi}}.
\label{eq:EntropyBound}
\end{equation}
This shows that the growth of the number of central height field configurations is only enough to make the half-system entanglement entropy upper bounded by $N\ln{2}$. The same argument applies to the exact excited states, while a generic excited state would not be subjected to the same Schmidt decomposition of Eq.~\eqref{eq:SchmidtDecomp} and therefore have EE upper bounded by Lieb's residue entropy of square ice instead~\cite{Lieb67}. Recently, a 2D generalization of the Fredkin model was proposed on the hexagonal lattice with the dimer constraint~\cite{LukeEA22}. In Appendix~\ref{sec:Generalizations}, we define a similar 2D generalization of the Fredkin spin chain to the square lattice with the FPL constraint. By the logic of the upper bound above this model presumably also faces a ground state bipartite entanglement entropy bounded by area law. The same conclusion is expected to hold for other height-conserving dimer models considered recently~\cite{ZhengEA22a}.

From a tensor network point of view, the ground states of both models can be view as given by projected entangled paired state (PEPS), with virtual legs corresponding to our physical degrees of freedom, and no degrees of freedom on the physical legs. In the tensor network language, the EE scaling is given by the number of bonds crossed by the boundary between subsystems. So to achieve an area law breaking entanglement entropy or a real 2D generalization to the Fredkin model, one may consider constructing a 3D holographic tensor network for which the physical legs only live on the bottom layer, while for the rest of the layers, the virtual leg from the previous layer plays the role of a physical leg~\cite{RafaelEA19}.

%
%%
%%%
\section{Upper bound on the spectral gap}
\label{sec:Bulkgap}
%%%
%%
%
The classical six-vertex model at the homogeneous point, at which the Boltzmann weights of the $a$-, $b$-, and $c$-type vertices of Fig.~\ref{fig:HeightRep} (a) are equal, is critical~\cite{Baxter16}, with algebraically decaying bond-bond correlators. By construction, the ground state of our quantum Hamiltonian has the same equal-time correlation function as the classical model, which mean the spectral gap between ground state and first excited state must vanish in the thermodynamic limit~\cite{ColomoCorr, ColomoCorr1, Dubail, BelovEA20}. In this section, we construct a trial state (Eq.~\eqref{eq:ExcitedTrialState}) to show that the excitation gap is upper bounded by a quantity that decays exponentially with system size.

\begin{figure}[t!bh]
	\centering
	\includegraphics[width=0.7\linewidth]{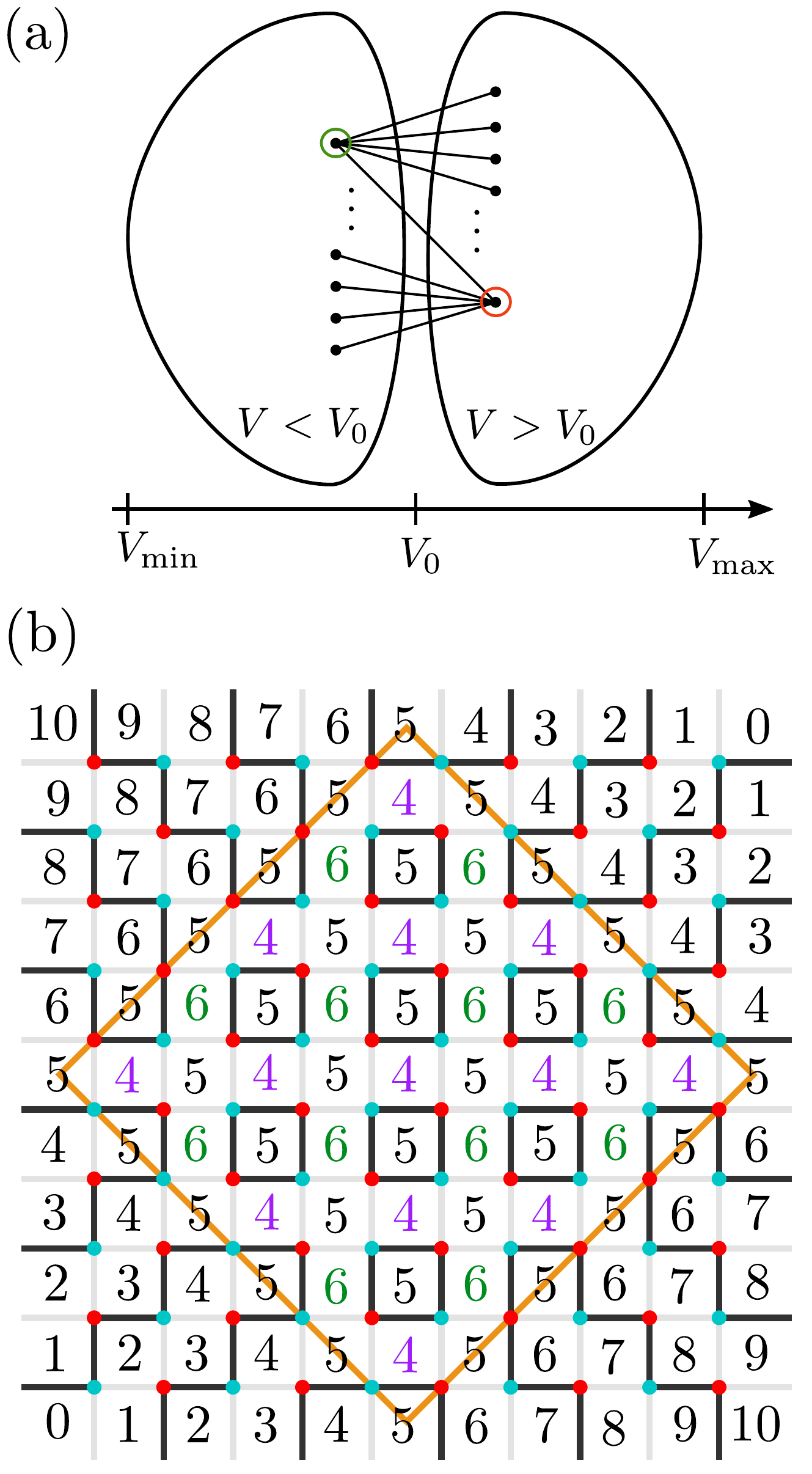} 
	\caption{(a) Fully packed loop configurations with domain-wall boundary conditions partitioned into two sets, dictated by the volume of the configuration. Only configurations in the vicinity of the boundary between the two sets, as measured in a volume metric, contribute to the energy expectation of the trial state in Eq.~\eqref{eq:ExcitedTrialState}. (b) One of the $\binom{M}{\frac{M+1}{2}}$ configurations (with $M = N^2/4$) with volume $V_0 - 1$. The other configurations are obtained by simultaneously flipping an equal number of plaquettes with height $6$ and $4$.}
	\label{fig:Gapless}
\end{figure}

The trail state needs to meet two requirements: it has to be both orthogonal to the ground state, and the Hamiltonian must have an expectation value that approaches zero in thermodynamic limit for it to provide an upper bound on the excitation gap. The first requirement can be satisfied by adding a ``twist'' to the ground state superposition~\cite{BravyiEA12}, such that when taking the inner product with the ground state, the two parts with opposite signs cancel. The second requirement calls for a carefully chosen boundary between the two sets of configurations, such that the Hamiltonian annihilates the intra-set contributions, leaving only contributions from the interface of the sets in the energy expectation value. A convenient choice of the set boundary is halfway between the highest $V_{\mathrm{max}} = \frac{1}{3}N(N+1)(2N+1)$ and lowest $V_{\mathrm{min}} = \frac{1}{3}N(N+1)(N+2)$ volume configurations (as found by summing up the configurations in Eq.~\eqref{eq:ExtremalHeight}). The two sets contain the same number of configurations, and one has to traverse one of the many such configurations to go from configurations with a volume smaller than their average, $V_0 \coloneqq \frac12(V_{\mathrm{min}} + V_{\mathrm{max}}) = \frac{N(N+1)^2}{2}$, to one with a larger volume, see Fig.~\ref{fig:Gapless}. We thus pick the trail state
\begin{equation}
     \ket{\pi} = \sum_{\pazocal{F} \in \mathrm{FPL\ with\ DWBC1}\\ }\mathrm{sgn}(V(\pazocal{F})-V_0)  \ket{\pazocal{F}}.
\label{eq:ExcitedTrialState}
\end{equation}
It immediately follows that $\braket{\mathrm{GS}}{\pi}=0$ and $\braket{\pi}{\pi} = A(N)$. Before evaluating its energy expectation value, we first count the numbers of configurations near the volume interface $V_0$. A representative configuration with volume $V_0 - 1$ is shown in Fig.~\ref{fig:Gapless}. The flippable plaquettes of this configuration are all located inside a diamond with half the size of the lattice. Outside this diamond the plaquettes are frozen. Inside, every other plaquette is flippable, giving a total number of $M \coloneqq \frac{N^2}{4}$ flippable plaquettes. Among these there are $\frac{M+1}{2}$ plaquettes with height $\frac{N}{2}-1$ and $\frac{M-1}{2}$ with height $\frac{N}{2}+1$~\footnote{We have assumed that $N/2$ is odd. For $N/2$ being even, the construction is analogous with slight modifications.}. The number of configurations with volume $V_0-1$ can thus be enumerated as
\begin{equation}
    \sum_{n=0}^{\frac{M-1}{2}} \binom{\frac{M-1}{2}}{n} \binom{\frac{M+1}{2}}{n} = \binom{M}{\frac{M+1}{2}},
\end{equation}
since one can simultaneously flip any number of pairs of plaquettes of heights $\frac{N}{2}-1$ and $\frac{N}{2}+1$ from this reference state and remain in vicinity to the volume boundary. Each of these configurations can brought across the volume boundary by $\frac{M+1}{2}$ Hamiltonian terms flipping one of the plaquettes with height $\frac{N}{2}- 1$. We have
\begin{equation}
\begin{aligned}
    \bra{\pi}H\ket{\pi}=&\sum_{V(\pazocal{F}),V(\pazocal{F}')=V_0\pm 1 } \bra{\pazocal{F}}H\ket{\pazocal{F}'}\\
    =&\frac{M+1}{2} \binom{M}{\frac{M+1}{2}}.
\end{aligned}
\label{eq:PiExpectation}
\end{equation}
Using the asymptotic behaviour of $A(N)$ form Eq.~\eqref{eq:ASM}~\cite{Note20} we find
\begin{equation}
     \frac{\bra{\pi}H\ket{\pi}}{\bra{\pi} \ket{\pi} } \sim N^{\frac{41}{36}} \left(\frac{\sqrt[4]{2}}{3\sqrt{3}/4}\right)^{N^2}\to 0,
\end{equation}
proving that the Hamiltonian is gapless in the thermodynamic limit. A remark is in order here to point out the connection between our proof of gaplessness and the ``arctic curves'' of the six-vertex model~\cite{ArcticCircle}. The vanishing asymptotics above strongly relies on the fact that near $V_0$, the region containing flippable plaquettes only occupy half of the lattice and the corners outside the orange rhomboid in Fig.~\ref{fig:Gapless} (b) are frozen.

%

%
%%
%%%
\section{Conclusions}
\label{sec:Concl}
%%%
%%
%
We constructed a quantum fully packed loop model with a Rokhsar--Kivelson type Hamiltonian~\cite{RokhsarKivelson88}, in which configurations permit multiple equivalent formulations from the classical statistical mechanics and combinatorics literature.
By making the model frustration-free the quantum ground state becomes an equal superposition of configurations from the classical space of configuration. We showed that the bulk configurations are heavily constrained by the boundary, to the point that certain boundary configurations imply product eigenstates. The bulk Hamiltonian is not ergodic in the entire Hilbert space, but only within each Krylov subspace, as dictated by the boundary configuration.  Each ergodic subspace has its own lowest energy eigenstate, which are equidistant in energy across subspaces. Owing to enumerable half-system configurations by recursion, we performed an exact lattice calculation of the ground state bipartite entanglement entropy for systems of sizes up to $18 \times 18$ giving area law scaling.

Our methodology may turn useful in the study of other height models and in related studies of ergodicity breakdown induced by boundary terms. One possible generalization is to consider a $\mathbb{Z}_n$ generalized model involving a boundary condition that alternates with period $n$ instead of two. One may also expect the emergence of interesting phases and refined structures in the ergodicity breaking for fully packed loop models adopted to non-bipartite lattices. Lattices of interest include the triangular one~\cite{MoessnerEA01, FendleyEA02, ZhengEA22b}, or more exotic ones such as the Kagome lattice~\cite{MisguichEA02}, or even aperiodic tilings like the Penrose~\cite{FlickerEA20} and the Amman--Beenker tiling.

One could attempt to construct a microscopic Hamiltonian that makes the FPL constraint emerge, like what was done for the dimer model in Refs.~\onlinecite{CanoFendley, Klein} using Klein terms of the Hamiltonian. There are also alternative ways to implement the DWBC, for example, by employing an antiferromagnetic interaction along the boundary. The outcome of this choice, other than making the ground state two-fold degenerate and the exact excited states reordered in energy, is that the entire Hamiltonian will have a $\mathbb{Z}_2$ symmetry.

One can also introduce dynamic terms in the boundary Hamiltonian, so that the fragmentation is removed and the unique ground state becomes the superposition of FPLs with all boundary configurations. It would be of interest to explore the consequences of that on the EE scaling. In addition, it is also interesting to think of whether the Hamiltonian can be modified to reallocate the exact excited states to mid-spectrum QMBS states. \\

\begin{acknowledgments}
ZZ thanks Filippo Colomo, Hosho Katsura, Israel Klich, Yuan Miao and Jeffrey Teo for fruitful discussions. ZZ acknowledges the kind hospitality of the workshop ``Ramdomness, Integrability and Universality'' at the Galileo Galilei Institute for Theoretical Physics during the final stage of this work. HSR acknowledges helpful conversations with Michele Burrello, Paul Fendley, Thomas Scaffidi, Steven H. Simon, Felix Flicker, Olav F.~Sylju{\aa}sen, and Weronika Wrzos-Kaminska. ZZ was supported in part by the European Research Council under grant
742104. HSR was supported by a research grant (40509) from VILLUM FONDEN. HSR was also supported by VILLUM FONDEN via the Centre of Excellence for Dirac Materials (Grant No.~11744). We acknowledge the computing resource norlx68 at Nordita. Nordita is partially supported by Nordforsk. 

\end{acknowledgments}

\bibliography{FPLs}

\begin{appendix}
\onecolumngrid

%
%%
%%%
\section{Proof of ergodicity of plaquette flipping within boundary sectors}
\label{sec:Ergodicity}
%%%
%%
%
In this appendix we prove by induction that each Krylov subspace specified by the boundary configuration ($H_{\partial}$) is ergodic with respect to the bulk Hamiltonian ($H_{\mathrm{bulk}}$). 

For a $2\times 2$ lattice, there are at most two choices of the bulk configuration, differing by the height on the central plaquette, for any fixed boundary configuration. These two are related by the flipping the central plaquette. 

\begin{figure}[t!bh]
	\centering
	\includegraphics[width=0.35\linewidth]{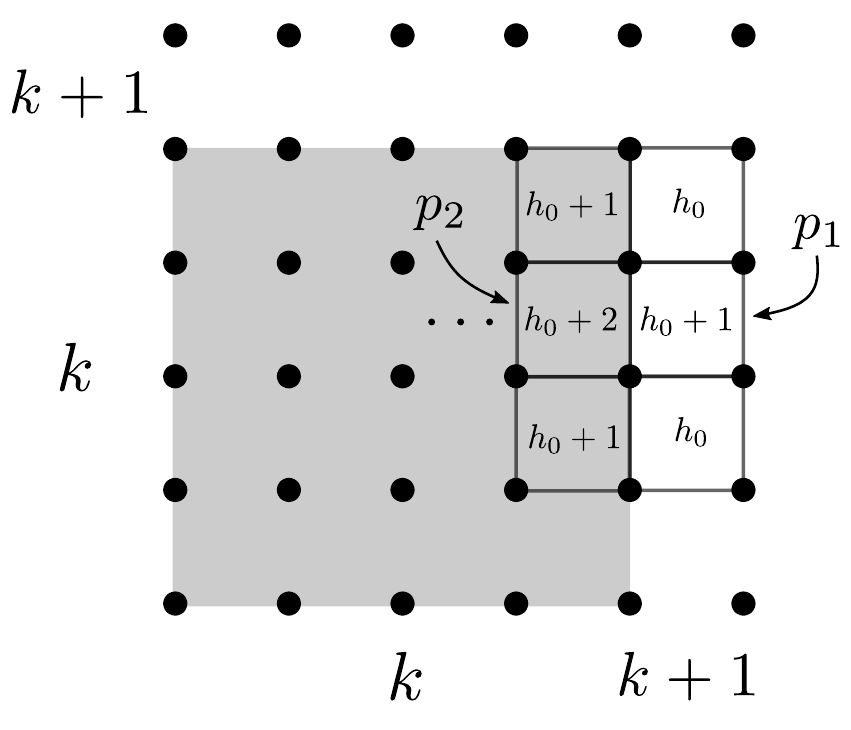} 
	\caption{Proving ergodicity of the bulk Hamiltonian by induction: Two configurations differing only by the height of one plaquette along the strip added in step $(k+1)$ are either related by flipping that plaquette ($p_1$), or if it is not flippable, and a sequence of flipping operations in the $k\times k$ bulk (by its proven ergodicity) will make it flippable. Here, the height configuration of one of the two configurations differing at $p_1$ and $p_2$ is shown, the other configuration has height $h_0-1$ on $p_1$ and $h_0$ on $p_2$.}
	\label{fig:Induction}
\end{figure}

Suppose we have proven the ergodicity for (the interior of) a lattice of size $k \times k$ for any fixed boundary configuration, that is, if such a boundary allows multiple bulk configurations, they can all be transformed into each other by sequences of plaquette flips. Then for a lattice of size $(k+1) \times (k+1)$, for which the bottom left corner is a $k \times k$ lattice with established ergodicity, see Fig.~\ref{fig:Induction}, we just need to show that any configuration of the newly added strip above the top and to the right with identical bulk configurations, transform into each other by sequences of plaquette flips.

To show this, we only need to show that sequences of plaquette flips can transform any two strip configurations differing only by the height of any single plaquette into each other, because once that is established, configurations differing at multiple places can be connected step by step transitively. If the difference is located in the corner of the strip, then the two configuration are related by simply flipping the corner plaquette. Otherwise, if the heights $h_0\pm 1$ differ on a plaquette $p_1$, see Fig.~\ref{fig:Induction}, on the right boundary, its neighbors above and below must both have height $h_0$, or there are at least two plaquettes along the strip with different heights between the two strip configurations in question, contradicting our assumption.

All we need to do now is to show that there is another bulk configuration in the $k\times k$ lattice compatible with the strip configuration on its boundary, for which the adjacent plaquette $p_2$ to the left of $p_1$ has height $h_0$, between the two plaquettes of the strip with height $h_0\pm 1$. Then these two configurations of the $k+1\times k+1$ lattice with identical heights in the $k\times k$ lattice are related by flipping plaquette $p_1$. And this must be the case, because if not, $p_2$ must have $h_0+2$ (resp.~$h_0-2$) if the height on $p_1$ is $h_0+1$ (resp.~$h_0-1$), with neighbors above and below both of same height as $p_1$, otherwise, the flippability of $p_2$ implies that there is another bulk configuration with height $h_0$ on $p_2$ making $p_1$ flippable. This argument goes on until we reach at the left boundary, with a monotonic gradient along this row, see Fig.~\ref{fig:Induction}. As we established in the exhaustion of product eigenstates, this means the entire lattice must be covered by horizontal straight gradient lines, making the ergodicity proof irrelevant for the boundary configuration under discussion.

In the periodic boundary case, ergodicity was proven in Ref.~\onlinecite{HermeleEA04} for the ``zero-flux'' sector, which contains two configurations with every plaquette flippable, called ``ideal states'', and they are connected to each other by flipping half of the plaquettes. Since a global extremum implies a local extremum, each configuration has at least two flippable plaquettes which can be flipped to reduce the difference between maximum and minimum heights until one reaches one of the two ideal states. As elegant as this proof is, it does not work for fixed boundary conditions since the extrema can be located on the boundary. Modulo exceptions of configurations with no flippable plaquettes, which we provide the precise criteria for in Sec.~\ref{sec:Fragmentation}, the ergodicity in other Krylov subspaces under periodic boundary conditions can be proved without too many modifications of the above proof.

In Ref.~\onlinecite{Kondakor21} it is stated that the ``topological sectors'' are classified by the height change along two perpendicular non-contractible loop directions in the case of periodic boundary conditions. There are $2N+1$ such sectors and the bulk Hamiltonian is ergodic within each of these. This can be seen by cutting the torus along two orthogonal non-contractible loops, due to periodicity, the boundary configuration are uniquely determined by the height along these two loops, which form an ``L"-shaped strip. The kinetic terms in the Hamiltonian can change any configuration along the strip to any other with the same magnetization of arrows pointing in and out. The disconnected sectors can simply be enumerated with a representative where arrows pointing in come before all arrows pointing out, from each magnetization. The number of possible domain wall locations is $2N+1$ for $2N$ spins. Note that this is different from the the case of open boundary conditions considered above, for which the number of sectors grows exponentially with system size. We caution, however, that the meaning of ``topological sectors'' here is different from the more conventional usage, the latter in which there are $4^g$ disjoint sets of configurations for a genus $g$ graph that can not be transformed into each other by sequences of plaquette flips.

Finally, ergodicity of plaquette flipping moves among FPLs or dimer coverings is trivial on the hexagonal lattice, but less understood on the (tripartite) triangular lattice~\cite{MoessnerEA01, FendleyEA02}. 

\section{Sufficiency condition for exact excited product states}
\label{sec:sufficient}
\begin{figure}[t!bh]
	\centering
	\includegraphics[width=0.35\linewidth]{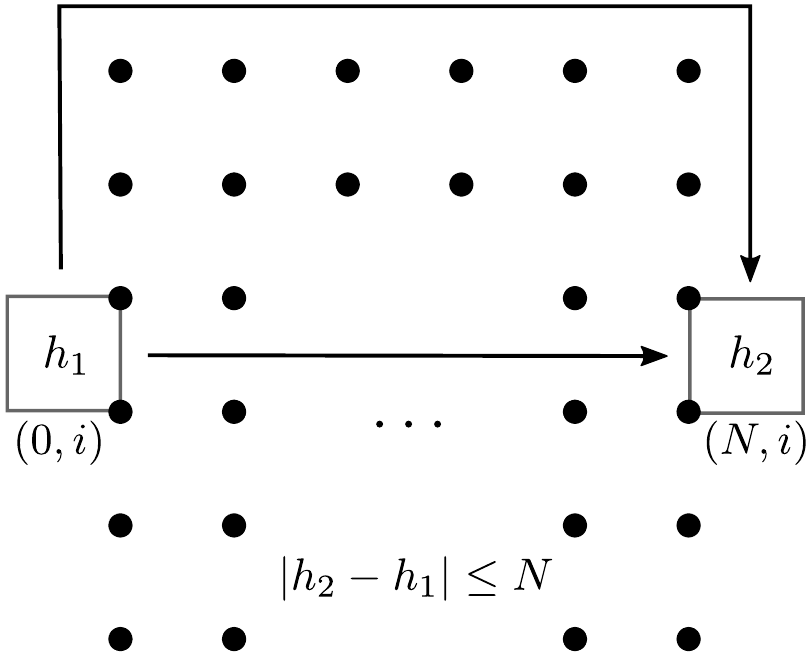} 
	\caption{An illustration of the second clause of the sufficiency condition for an exact excited state to be a product state: The height change going around two corners of the lattice cannot exceed the maximum possible height change of $N$, corresponding to going straight from one side to the opposite side.}
	\label{fig:BackwardsInduction}
\end{figure}

The fact of zero net inflow of arrows in the six-vertex language for all subsystems of the lattice is sufficient to guarantee exact excited states is proven by construction. We start with an $N\times N$ lattice configuration of which the boundary satisfies this condition, and show how to construct all the configurations of the $(N-1)\times (N-1)$ interior layer that satisfy the same condition. By backwards induction the procedure keeps going until a lattice with only one plaquette is reached, which can be easily shown to allow a consistent configuration by examination. In practice, we use backwards induction in the height degree of freedom. An illustration of the second clause of the sufficiency condition is shown in Fig.~\ref{fig:BackwardsInduction}.

A boundary configuration for a system of size $N$ satisfying the condition guarantees that heights of plaquettes on opposite sides of the same row or column differ by at most $N$. We imagine filling in the heights on the plaquettes in the next layer one by one, starting from the plaquette with coordinate $(1,1)$. If its two neighbors on the left and below have different heights, then its height is fixed to be the average of those two. Otherwise, one still needs to check if the height difference between plaquette pairs $(1,0)$, $(1, N)$, and $(0,1)$, $(N,1)$ is exactly $\pm N$. If so the height of $(1,1)$ plaquette must be increase (resp. decrease) by one from its neighbors on the left and below. One can easily check that if the two requirements are simultaneously met one can never run into a contradiction. In all other cases we have a freedom in choosing the height. The algorithm devised in Appendix~\ref{sec:MonteCarlo} makes use of these observations. Next, we fill in the height of cell $(1,2)$ in the same way, only this time checking the horizontal constraint of whether height at $(1,1)$ and $(1,N)$ differ exactly by $N-1$. After finishing the first row, we fill in the rest of the first column in the same way. Then we proceed to the second last row, checking the height differences at $(1, i)$ and $(N, i)$ instead. And finally, we fill the second last column, checking differences in heights at plaquette $(i,1)$ and $(i, N)$. In the process, we have guaranteed that the boundary configuration of the $(N-1)\times (N-1)$ lattice satisfies the same condition. The induction goes on until the $3\times 3$ lattice, for which we know that the two ends of the middle row and column can differ by at most two, so we can always fill in the height in the center. The resulting height distribution from each outcome of the above procedure would map to a unique FPL configuration, and the exact excited state corresponding to their common boundary configuration is given by the uniform superposition of all of them.

%
%%
%%%
\section{Enumeration of half-system configurations}
\label{sec:Recursive}
\begin{figure}[t!bh]
	\centering
	\includegraphics[width=0.8\linewidth]{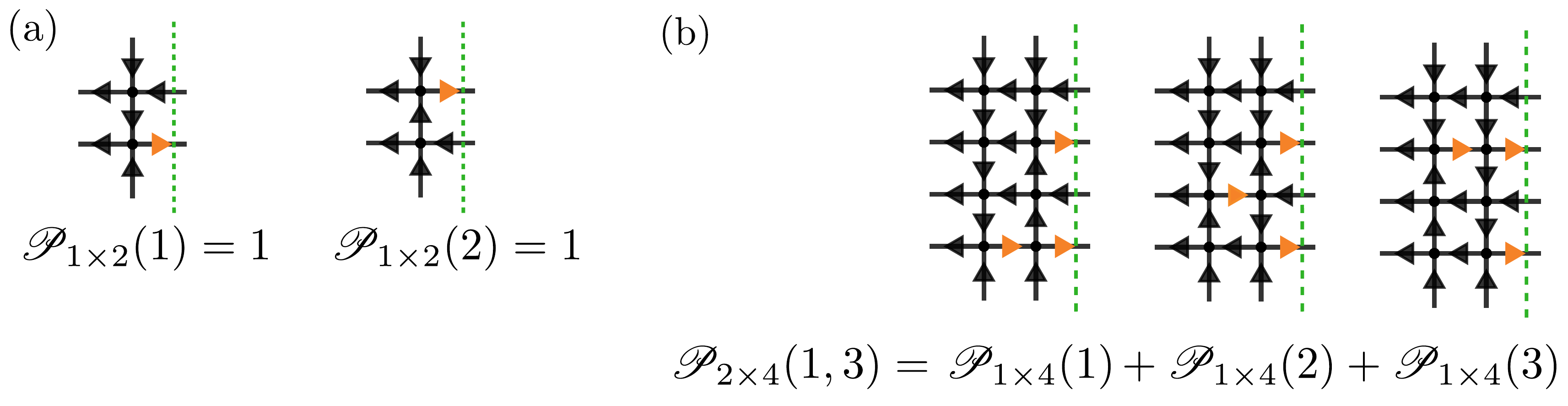} 
	\caption{(a) The two vertex configurations of size $1 \times 2$ with kinks of vertical arrows on the first and second row. (b) An illustration of recursively enumerating the configurations of size $2\times 4$ with reversed horizontal arrows on the right boundary on the first and third row, from the counting of configurations of size $1 \times 4$. The orange arrows indicate how each closest pair of right arrows must vertically sandwich a right arrow on their left. The general recursion relation is given by Eq.~\eqref{eq:RecursionRel}.}
	\label{fig:RecursionRelation}
\end{figure}

%%%
%%
%
The recurrence relation for enumerating half-system FPL configurations can be explained in the language of the six vertices (top row of Fig.~\ref{fig:HeightRep} (a))~\footnote{This recurrence relation was derived in Ref.~\onlinecite{Nianqiao16} where the author proved the relation after mapping to equivalent Gelfand--Tsetlin patterns. Here we give a proof directly from the six-vertex degrees of freedom.}. In this language the height change going one step up increases (resp.~decreases) by one if a right (resp.~left) pointing arrow is crossed. Since the domain wall boundary condition fixes the height at the top and bottom of the middle column to be the same, the height profile along the middle cut is a Dyck walk~\cite{Salberger2EA17, ZhangKlich17}, therefore having half of the arrows pointing left and half pointing right. Along the zeroth column on the left boundary, all arrows are pointing left. If we label the locations of the arrows that are now pointing right instead along the $\frac{N}{2}$'th column with a vector $\vec{x} = (x_1, x_2,\dots, x_{N/2} )$, then $\lvert \pazocal{P}_{L,\vec{m}}(N/2) \rvert$ in Eq.~\eqref{eq:SchmidtDecomp} becomes a function of $\vec{x}$, which we define to be $\mathscr{P}_{N/2\times N}(\vec{x})$:
\begin{equation}
    \mathscr{P}_{N/2\times N}(\vec{x}) \coloneqq \lvert \pazocal{P}_{L, \vec{m}}(N/2) \rvert,
    \label{eq:Recursion}
\end{equation}
where $x_i$ denotes the location of the $i$'th arrow (counting from bottom to top) reversed to point right in this column, and ${N/2\times N}$ specifies the dimension of the half-system. 
For $N=2$, there is only one allowed configuration for each boundary condition, so $\mathscr{P}_{1\times 2}(1)=\mathscr{P}_{1\times 2}(2)=1$, see Fig.~\ref{fig:RecursionRelation} (a). While we only care about right boundary configurations with half of the arrows flipped in the end, in order to enumerate them recursively, we also need the number of configuration with less than half of the arrows flipped. In fact, we also have $\mathscr{P}_{1\times N}(i) = 1, \forall~1 \le i \le N$, corresponding to a kink of vertical arrows on the $i$'th row. 
On the left half-lattice, all the horizontal arrows point to the left along the $0$'th column. After taking into account the two inward vertical arrows in the first column at the lower and upper boundary, only one horizontal arrow changes to the right, as there can only be one kink along the vertical line (any more than that would violate the six-vertex rules at some point). Once a horizontal arrow is reversed along the first column, two things can happen after the next column: either it will be reversed to point left again, or one more arrow will be flipped to point right. Only the latter case is relevant to us because, otherwise, we can not accumulate a number of reversed arrows equal to the number of columns $N/2$, which is half of the number of rows, see Fig.~\ref{fig:RecursionRelation} (b).

A recurrence relation can be summarized from the simple fact that the right arrows along the previous column must be evenly sandwiched between the rows of each adjacent right arrow pairs in the next column. If instead, all the horizontal arrows between any two closest right arrows in the next column (including upper and lower boundaries) were pointing to the left, then we would have either an even number of kinks of vertical arrows, or a violation of the six-vertex rules, neither of which is allowed. We have:
\begin{equation}
    \mathscr{P}_{(\frac{N}{2}+1)\times (N+2)}(\vec{x})=\sum_{\substack{x_1\le y_1\le x_2 \le y_2 \le\ldots \le y_{\frac{N}{2}}\le x_{\frac{N}{2}+1}\\ y_1\ne y_2\ne \ldots \ne y_{\frac{N}{2}}}}\mathscr{P}_{\frac{N}{2} \times (N+2)}(\vec{y}),
    \label{eq:RecursionRel}
\end{equation}
which states that between each neighbouring pairs of reversed horizontal arrows in the current column, there must be one in the previous column, and they must all be distinct.

%
%%
%%%
\section{Monte Carlo method}
\label{sec:MonteCarlo}
%%%
%%
%
Here we devise a Monte Carlo method to verify the recursion relation of Appendix~\ref{sec:Recursive}. Given a protocol to generate central height configurations $\vec{m}$ uniformly, the principle is to approximate Eq.~\eqref{eq:Entropy} as 
\begin{equation}
S = -\sum_{\lbrace \vec{m} \rbrace} p_{\vec{m}} \ln{ p_{\vec{m}} } \approx -\sum_{\lbrace \vec{m} \rbrace_{\mathrm{MC}} } \tilde{p}_{\vec{m}} \ln{ \tilde{p}_{\vec{m} } },
\label{eq:MCprinciple}
\end{equation}
where the distribution $\tilde{p}_{\vec{m}}$ approximates Eq.~\eqref{eq:pmcoeff} as
\begin{equation}
\tilde{p}_{\vec{m}} = \frac{M_{\vec{m}} M_{\vec{N}-\vec{m}}}{M_{\mathrm{MC}}},
\label{eq:ApproximateDist}
\end{equation}
with $M_{\vec{m}}$ being the number of times the central height $\vec{m}$ is sampled, $M_{\mathrm{MC}}$ is the total number of configurations sampled, and where $\lbrace \vec{m} \rbrace_{\mathrm{MC}} $ is the set of distinct heights $\vec{m}$ encountered in the $M_{\mathrm{MC}}$ samples.

The key part of the algorithm is to uniformly generate the central heights $\vec{m}$. To achieve this, we iterate row by row from the lower left corner of the graph (with DWBC1 imposed) and draw height configurations with the following rule:
\begin{equation}
h(i,j) = \begin{cases}
\frac{1}{2}\left[ h(i-1,j) + h(i,j-1) \right] & \mathrm{if}~ h(i-1,j)\neq h(i,j-1) \\
h(i-1,j)-1  & \mathrm{if}~ h(i-1,j)= h(i,j-1)~\mathrm{and}~h(i-1,j)+1>h_{\mathrm{max}}(i,j)  \\
h(i-1,j)+1  & \mathrm{if}~ h(i-1,j)= h(i,j-1)~\mathrm{and}~h(i-1,j)-1<h_{\mathrm{min}}(i,j)  \\
h(i-1,j)+1~\mathrm{with~prob.~}\frac{1}{2}  & \mathrm{if}~ h(i-1,j)= h(i,j-1)~\mathrm{and}~h_{\mathrm{min}}(i,j) < h(i-1,j)<h_{\mathrm{max}}(i,j)  \\
h(i-1,j)-1~\mathrm{with~prob.~}\frac{1}{2}  & \mathrm{if}~ h(i-1,j)= h(i,j-1)~\mathrm{and}~h_{\mathrm{min}}(i,j) < h(i-1,j)<h_{\mathrm{max}}(i,j)
\end{cases},
\label{eq:HeightRule}
\end{equation}
where we invoke the globally extremal height configurations, which due to the imposed domain-wall boundary conditions take the form
\begin{equation}
h_{\mathrm{min}}(i,j) = \lvert i - j \rvert, \hspace{20pt} \mathrm{and} \hspace{20pt} h_{\mathrm{max}}(i,j) = \begin{cases}
i+j & \mathrm{if}~i+j \le N \\
2N - i - j & \mathrm{if}~i+j > N
\end{cases},
\label{eq:ExtremalHeight}
\end{equation}
with $h_{\mathrm{max}}(i,j)$ being shown in Fig.~\ref{fig:HeightRep} (c). To make sure all height configurations are generated with equal probability we trace the number of random numbers drawn to achieve the lower two lines of Eq.~\eqref{eq:HeightRule}, $R$, and weigh the corresponding branch in the configuration binary tree by $2^R$.

In practice the main hurdle to overcome with this algorithm is to make $M_{\mathrm{MC}}$ large enough to guarantee that all the $\pazocal{N}_{\vec{m}} \sim 2^N/\sqrt{\pi N /2}$ (Eq.~\eqref{eq:Nomvecs})) central heights are encountered with sufficient multiplicity, otherwise accuracy is expected drop and the entropy is eventually underestimated. For instance, with $N = 12$, a single run with $M_{\mathrm{MC}} = 2\cdot 10^9$ was enough to make $\lvert \lbrace \vec{m} \rbrace_{\mathrm{MC}} \rvert = 922 < 924 =  \pazocal{N}_{\vec{m}} $.

%
%%
%%%
\section{A less constrained frustration-free 2D model}
\label{sec:Generalizations}
%%%
%%
%
One can relax the fixed boundary height and impose the Fredkin moves $\ket{\uparrow\uparrow\downarrow} \leftrightarrow \ket{ \uparrow\downarrow\uparrow}$ and $\ket{\downarrow \uparrow \downarrow } \leftrightarrow \ket{\uparrow\downarrow\downarrow}$ in the vertical and horizontal direction by constructing a Hamiltonian made out of projectors onto singlets of Fredkin-flipped states in both the vertical and horizontal direction. Sticking to the bipartite rules of Fig.~\ref{fig:HeightRep} (a) and (b) we must distinguish between even and odd plaquettes in doing so, resulting in the Hamiltonian
\begin{align}
H &= \sum_{p\in\lbrace \mathrm{o}, \mathrm{e} \rbrace } \Big[ \ket{E_{p}}\bra{E_{p}} + \ket{W_{p}}\bra{W_{p}} + \ket{N_{p}}\bra{N_{p}} + \ket{S_{p}}\bra{S_{p}} \Big], \label{eq:Hamiltonian2} \\
\ket{E_{\mathrm{o}}} &= \ket{\Eastoa} - \ket{\Eastob},~\ket{E_{\mathrm{e}}} = \ket{\Eastea} - \ket{\Easteb}, \label{eq:East} \\
\ket{W_{\mathrm{o}}} &= \ket{\Westoa} - \ket{\Westob},~\ket{W_{\mathrm{e}}} = \ket{\Westea} - \ket{\Westeb}, \label{eq:West} \\
\ket{N_{\mathrm{o}}} &= \ket{\Northoa} - \ket{\Northob},~\ket{N_{\mathrm{e}}} = \ket{\Northea} - \ket{\Northeb}, \label{eq:North} \\
\ket{S_{\mathrm{o}}} &= \ket{\Southoa} - \ket{\Southob},~\ket{S_{\mathrm{e}}} = \ket{\Southea} - \ket{\Southeb}, \label{eq:South} 
\end{align}

\begin{figure}[t!bh]
	\centering
	\includegraphics[width=0.5\linewidth]{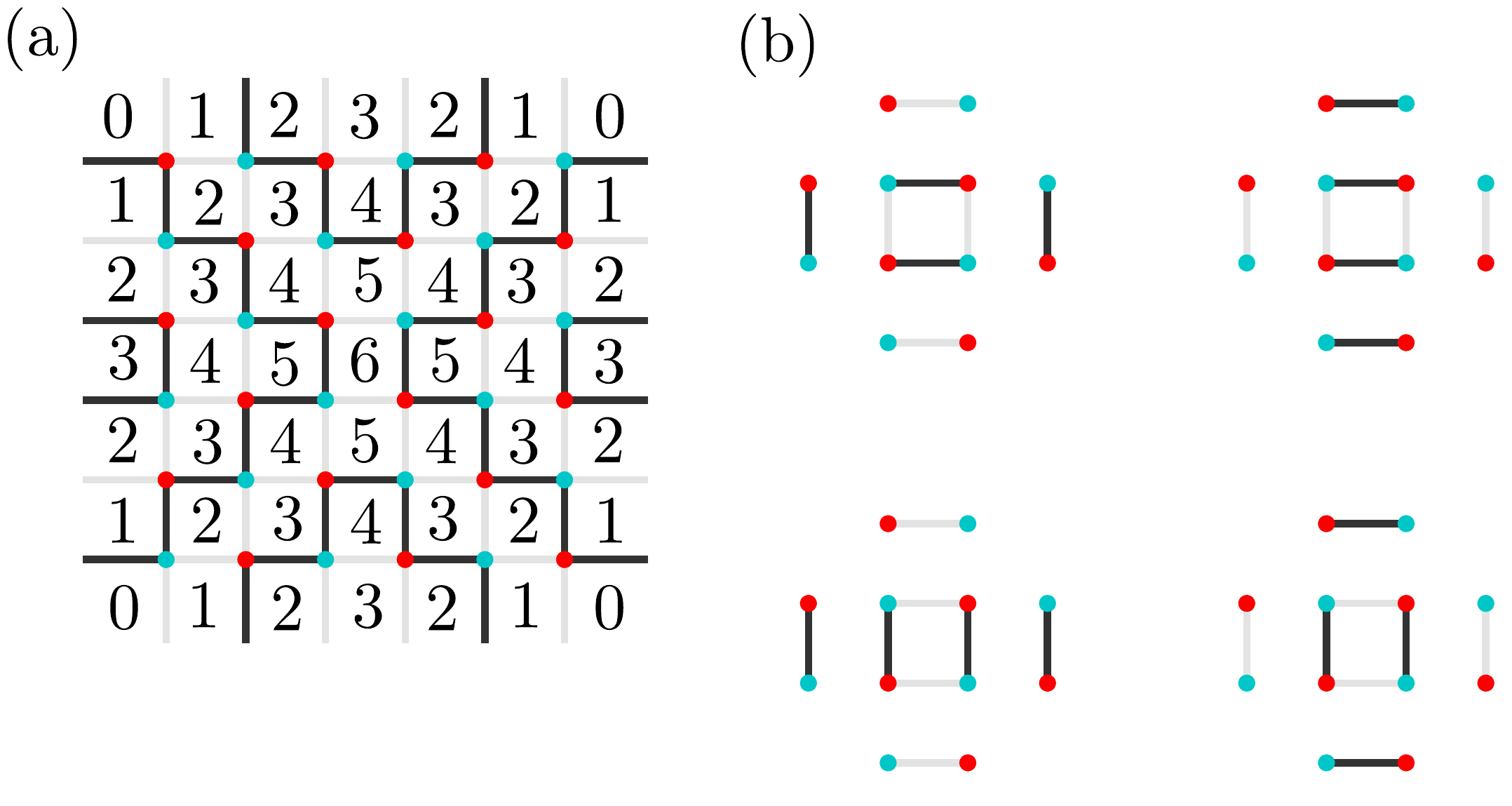} 
	\caption{(a) Maximal height configuration for the model defined in Eq.~\eqref{eq:Hamiltonian2}. (b) The four plaquettes that are not flippable in the model of Eq.~\eqref{eq:Hamiltonian2} but that are flippable in the quantum dimer model.}
	\label{fig:HeightConfig2}
\end{figure}

In Fig.~\ref{fig:HeightConfig2} panel (a) we show the maximal height configuration of this model, and in panel (b) we show the plaquettes that are not flippable in this model, to ensure a positive height.
For system sizes of $N = 4,~6$ we confirmed by explicit enumeration~\footnote{The number of configurations in the ground state superposition is $690$ (cf.~$42$ for the model in Sec.~\ref{sec:Model}) and $14\;058\;234$ (cf.~$7\;436$ for the model in Sec.~\ref{sec:Model}) for $N = 4$ and $N = 6$, respectively.} that this model exhibits a larger ground state bipartite entanglement entropy than the model of Sec.~\ref{sec:Model}. In fact, this is the case for any straight horizontal or vertical system partition. And even though this Hamiltonian poses as a perhaps more direct 2D generalization of the 1D Fredkin chain than Eq.~\eqref{eq:RKmodel}, it comes at the price of losing the combinatorial counting technology developed for FPLs with DWBC. And crucially, the simple bound of Eq.~\eqref{eq:EntropyBound}, or more precisely the growth of the number of height configurations, makes it immediately clear that also this model is bounded by area law bipartite entanglement entropy scaling in the ground state. 

\end{appendix}
\end{document}